%% file: sellwood.tex
\input mymacro

\linespacingstep=1 \multiplelines \normalbaselines
\ifnum\linespacingstep=1
  \footline{\hfil} 
  \headline{\ifnum\pageno=1
    \hfil
Revised version to appear in {\it The Astrophysical Journal\/}
% Submitted to {\it The Astrophysical Journal\/}
% Draft on \today
  \else
    {\headfont Axisymmetric bending modes} \hfil \tenrm\folio \hfil 
  Sellwood %\today
  \fi}
\else
   \hsize=16cm
   \vsize=21cm
   \raggedright
   \hyphenpenalty=10000
   \tolerance=5000
\fi
\raggedbottom

\def\mk{m_{\rm K}}
\def\kappaz{\kappa_{\rm z}}

\newbox\topbox
\long\def\mtopbox#1{\setbox\topbox=\vbox{#1}
  \topinsert{\box\topbox}\endinsert}

\mtopbox{\ifnum\linespacingstep=1 \vskip 0.5cm \else \vskip 5cm \fi
\centerline{\uppercase{\titlefont Axisymmetric bending oscillations of stellar disks}}}

\ifnum\linespacingstep=1 \blankline \else \vskip 2cm \fi
\centerline{\authorfont J. A. Sellwood
}
\vskip-3pt
\centerline{\affilfont Department of Physics and Astronomy, Rutgers University, Piscataway, NJ 08855}
\vskip-3pt

\ifnum\linespacingstep=1
  \blankline
  \centerline{Rutgers Astrophysics Preprint No 164}
\fi

\ifnum\linespacingstep=1 \blankline \else \vfill\eject \fi

{\narrower
\centerline{ABSTRACT}
\vskip 3pt
Self-gravitating stellar disks with random motion support both exponentially growing and, in some cases, purely oscillatory axisymmetric bending modes, unlike their cold disk counterparts. 
% Past theoretical work on the global bending of disk galaxies has used the simplifying approximation that the orbits of all stars are circular in the mid-plane; the circular orbit model misses much of the important physics, and many of the frequently rehearsed arguments based on results from this limit are invalid. 
A razor-thin disk with even a very small degree of random motion in the plane is both unstable and possesses a discrete spectrum of neutral modes, irrespective of the sharpness of the edge.  Random motion normal to the disk plane has a stabilizing effect but at the same time allows bending waves to couple to the internal vibrations of the particles, which causes the formerly neutral modes to decay through Landau damping.

Focusing first on instabilities, I here determine the degree of random motion normal to the plane needed to suppress global, axisymmetric, bending instabilities in a family of self-gravitating disks.  As found previously, bending instabilities are suppressed only when the thickness exceeds that expected from a na\"\i ve local criterion when the degree of pressure support within the disk plane is comparable to, or exceeds, the support from rotation.  A modest disk thickness is adequate for the bending stability of most disk galaxies, except perhaps near their centers.

The discretization of the neutral spectrum in a zero-thickness disk is due to the existence of a turning point for bending waves in a warm disk, which is absent when the disk is cold.  When the disk is given a finite thickness, the discrete neutral modes generally become strongly damped through wave-particle interactions.  It is surprising therefore that I find some simulations of warm, stable disks can support (quasi-)neutral, large-scale, bending modes that decay very slowly, if at all.

\ifnum\linespacingstep=1 \blankline \else \vskip 1cm \fi
\noindent{\it Subject headings}: galaxies: evolution --- galaxies: kinematics and dynamics --- galaxies: structure --- celestial mechanics: stellar dynamics --- instabilities --- methods: numerical \par}

\ifnum\linespacingstep=1 \bigskip \else \vfill\eject \fi

\input sect1

\input sect2

\input sect3

\input sect4

\input sect5

\input sect6

\input sect7

\bigskip
I wish to thank Alar Toomre for helping to clarify the issues raised by this work, for his constructive criticism of the first draft of this paper and for sending his notes on many aspects of his extensive unpublished work on bending oscillations.  Conversations with David Merritt, Rob Nelson, Colin Norman and Scott Tremaine, and correspondence with Agris Kalnajs, have also been helpful.  The hospitality of CITA, where some this work was done, is gratefully acknowledged.  This work was supported by NSF grant AST 93/18617 and NASA Theory grant NAG 5-2803.

\sectcount=-1
\ifnum\linespacingstep=1 \bigskip \else \vfill\eject \fi
\input appendA
\ifnum\linespacingstep=1 \bigskip \else \vfill\eject \fi
\input appendB

\ifnum\linespacingstep=1 \bigskip \else \vfill\eject \fi
\input table1
\input table2
\input table3

\ifnum\linespacingstep=1 \bigskip \else \vfill\eject \fi
\input refs

\ifnum\linespacingstep=1 \bigskip \else \vfill\eject \fi
\input captions

\end

%% file: mymacro.tex
\def\blankline{\vskip \baselineskip}
\def\DF{{\tenrm DF}}

\def\half{{\leavevmode\kern.1em\raise.5ex\hbox{\the\scriptfont0 1}\kern-.1em
/\kern-.1em\lower.25ex\hbox{\the\scriptfont0 2}}} %exercise 11.6
\def\quarter{{\leavevmode\kern.1em\raise.5ex\hbox{\the\scriptfont0 1}\kern-.1em
/\kern-.1em\lower.25ex\hbox{\the\scriptfont0 4}}}

\def\spose#1{\hbox to 0pt{#1\hss}} % from Scott Tremaine
\def\ltsim{\mathrel{\spose{\lower 3pt\hbox{$\mathchar"218$}}
     \raise 2.0pt\hbox{$\mathchar"13C$}}}
\def\gtsim{\mathrel{\spose{\lower 3pt\hbox{$\mathchar"218$}}
     \raise 2.0pt\hbox{$\mathchar"13E$}}}
\def\gtlt{\mathrel{\spose{\lower 3pt\hbox{$\mathchar"13E$}}
     \raise 3pt\hbox{$\mathchar"13C$}}}

% POOR MAN'S BOLD
\def\pmb#1{\setbox0=\hbox{$#1$}%
  \kern-0.25em\copy0\kern-\wd0
  \kern.05em\copy0\kern-\wd0
%   \kern-0.025em\raise.0433em\box0 }
  \kern-0.025em\raise.0433em\box0}

\def\today{\count99=\day
           \ifnum\count99>20 \count98=\day
                             \divide\count98 by 10
                             \multiply\count98 by 10
                             \advance\count99 by -\count98 \fi
           \number\day\ifcase\count99 th\or st\or nd\or rd\else th\fi
           ~\ifcase\month none\or January\or February\or March\or April\or
                  May\or June\or July\or August\or September\or October\or
                  November\or December\fi
           ~\number\year}

\long\def\Ignore#1{\relax}%

\newdimen\digitwidth
\setbox0=\hbox{0}
\digitwidth=\wd0

\def\cf{{\it cf.}}
\def\eg{{\it e.g.}}

\def\etc{{\it etc.}}
\def\ie{{\it i.e.}}

% set up possibility of variable linespacing in increments of one third
\newcount\linespacingstep
\newcount\linespacing
% default is single
\linespacingstep=1
\def\multiplelines{\linespacing=\linespacingstep
   \advance\linespacing by 2
   \multiply\normalbaselineskip by \linespacing
   \advance\normalbaselineskip by 2pt % to ensure rounding up
   \divide\normalbaselineskip by 3}

% fonts
\font\fiverm=cmr5	\font\sixrm=cmr6	\font\sevenrm=cmr7
\font\eightrm=cmr8 	\font\ninerm=cmr9	\font\tenrm=cmr10
\font\twelverm=cmr12 	

\font\fivei=cmmi5 	\font\sixi=cmmi6 	\font\seveni=cmmi7
\font\eighti=cmmi8 	\font\ninei=cmmi9 	\font\teni=cmmi10
\font\twelvei=cmmi12

\font\fivesy=cmsy5  	\font\sixsy=cmsy6 	\font\sevensy=cmsy7
\font\eightsy=cmsy8 	\font\ninesy=cmsy9 	\font\tensy=cmsy10
\font\magnifiedtensy=cmsy10 at 12pt

\font\fivebf=cmbx5  	\font\sixbf=cmbx6 	\font\sevenbf=cmbx7
\font\eightbf=cmbx8 	\font\ninebf=cmbx9 	\font\tenbf=cmbx10
\font\twelvebf=cmbx12

	\font\eightit=cmti8 	\font\nineit=cmti9
\font\tenit=cmti10	\font\twelveit=cmti12

\font\eightsl=cmsl8 	\font\ninesl=cmsl9	\font\tensl=cmsl10
\font\twelvesl=cmsl12

\font\eighttt=cmtt8 	\font\ninett=cmtt9 	\font\tentt=cmtt10
\font\twelvett=cmtt12

\font\tenex=cmex10

% font size macros
\def\eightpoint{\def\rm{\fam0\eightrm}% switch to 8-point type
\textfont0=\eightrm\scriptfont0=\sixrm\scriptscriptfont0=\fiverm
\textfont1=\eighti\scriptfont1=\sixi\scriptscriptfont1=\fivei
\textfont2=\eightsy\scriptfont2=\sixsy\scriptscriptfont2=\fivesy
\textfont3=\tenex\scriptfont3=\tenex\scriptscriptfont3=\tenex
\textfont\itfam=\eightit\def\it{\fam\itfam\eightit}
\textfont\slfam=\eightsl\def\sl{\fam\slfam\eightsl}
\textfont\ttfam=\eighttt\def\tt{\fam\ttfam\eighttt}
\textfont\bffam=\eightbf\scriptfont\bffam=\sixbf
\scriptscriptfont\bffam=\fivebf\def\bf{\fam\bffam\eightbf}
% \tt \ttglue=.5em plus.25em minus.15em
\normalbaselineskip=9pt
\ifnum\linespacingstep>1\multiplelines\fi
\setbox\strutbox=\hbox{\vrule height7pt depth2pt width0pt}
\let\sc=\sixrm\let\big=\eightbig\normalbaselines\rm}

\def\ninepoint{\def\rm{\fam0\ninerm}% switch to 9-point type
\textfont0=\ninerm\scriptfont0=\sixrm\scriptscriptfont0=\fiverm
\textfont1=\ninei\scriptfont1=\sixi\scriptscriptfont1=\fivei
\textfont2=\ninesy\scriptfont2=\sixsy\scriptscriptfont2=\fivesy
\textfont3=\tenex\scriptfont3=\tenex\scriptscriptfont3=\tenex
\textfont\itfam=\nineit\def\it{\fam\itfam\nineit}
\textfont\slfam=\ninesl\def\sl{\fam\slfam\ninesl}
\textfont\ttfam=\ninett\def\tt{\fam\ttfam\ninett}
\textfont\bffam=\ninebf\scriptfont\bffam=\sixbf
\scriptscriptfont\bffam=\fivebf\def\bf{\fam\bffam\ninebf}
% \tt \ttglue=.5em plus.25em minus.15em
\normalbaselineskip=11pt
\ifnum\linespacingstep>1\multiplelines\fi
\setbox\strutbox=\hbox{\vrule height8pt depth3pt width0pt}
\let\sc=\sevenrm\let\big=\ninebig\normalbaselines\rm}

\def\tenpoint{\def\rm{\fam0\tenrm}% switch to 10-point type
\textfont0=\tenrm\scriptfont0=\sevenrm\scriptscriptfont0=\fiverm%
\textfont1=\teni\scriptfont1=\seveni\scriptscriptfont1=\fivei%
\textfont2=\tensy\scriptfont2=\sevensy\scriptscriptfont2=\fivesy%
\textfont3=\tenex\scriptfont3=\tenex\scriptscriptfont3=\tenex%
\textfont\itfam=\tenit\def\it{\fam\itfam\tenit}%
\textfont\slfam=\tensl\def\sl{\fam\slfam\tensl}%
\textfont\ttfam=\tentt\def\tt{\fam\ttfam\tentt}%
\textfont\bffam=\tenbf\scriptfont\bffam=\sevenbf%
\scriptscriptfont\bffam=\fivebf\def\bf{\fam\bffam\tenbf}%
% \tt \ttglue=.5em plus.25em minus.15em
\normalbaselineskip=12pt%
\ifnum\linespacingstep>1\multiplelines\fi
\setbox\strutbox=\hbox{\vrule height8.5pt depth3.5pt width0pt}%
\let\sc=\eightrm\let\big=\tenbig\normalbaselines\rm}

\def\twelvepoint{\def\rm{\fam0\twelverm}% switch to 12-point type
\textfont0=\twelverm\scriptfont0=\eightrm\scriptscriptfont0=\sixrm
\textfont1=\twelvei\scriptfont1=\eighti\scriptscriptfont1=\sixi
\textfont2=\magnifiedtensy\scriptfont2=\eightsy\scriptscriptfont2=\sixsy
\textfont3=\tenex\scriptfont3=\tenex\scriptscriptfont3=\tenex
\textfont\itfam=\twelveit\def\it{\fam\itfam\twelveit}
\textfont\slfam=\twelvesl\def\sl{\fam\slfam\twelvesl}
\textfont\ttfam=\twelvett\def\tt{\fam\ttfam\twelvett}
\textfont\bffam=\twelvebf\scriptfont\bffam=\eightbf
\scriptscriptfont\bffam=\sixbf\def\bf{\fam\bffam\twelvebf}
\tt % \ttglue=.5em plus.25em minus.15em
\normalbaselineskip=14pt
\ifnum\linespacingstep>1\multiplelines\fi
\setbox\strutbox=\hbox{\vrule height10pt depth5pt width0pt}
\let\sc=\eightrm\let\big=\twelvebig\normalbaselines\rm}

\twelvepoint

%\vsize=24.6 true cm \hsize=15.7 true cm %max for A4 paper are 28 x 19
\vsize=22.6 true cm \hsize=17 true cm %max for US paper are 25 x 19
\clubpenalty=5000
\widowpenalty=5000
%\raggedbottom \topskip 10pt %don't allow any extra flexibility for pagelength

\hsize=17.7cm
\vsize=23.9cm

\font\headfont=cmti12
\font\titlefont=cmr12
\font\authorfont=cmcsc10 at 12pt
\font\affilfont=cmr8
\font\lonefont=cmr10
\font\ltwofont=cmti12

\def\sect#1{\par\goodbreak\bigskip
\centerline{\nextsect\lonefont\uppercase{#1}}\nobreak\vskip 3pt \nobreak}

\def\subsect#1{\par\goodbreak\medskip
\centerline{\nextsub\ltwofont~#1}\nobreak}

\def\sectandsub#1#2{\par\goodbreak\bigskip
\centerline{\nextsect\lonefont\uppercase{#1}}\nobreak\vskip 1ex
\centerline{\nextsub\ltwofont~#2}\nobreak}

\newcount\sectcount
\newcount\subcount
\newcount\ssubcount
\def\nextsect{\ifnum\sectcount>-1 \global\advance\sectcount by 1
        \number\sectcount. \global\subcount=0 \fi}
\def\nextsub{\global\advance\subcount by 1 \number\sectcount.\number\subcount.
        \global\ssubcount=0}
%        \global\ssubcount=96}
\def\nextssub{\global\advance\ssubcount by 1
\number\sectcount.\number\subcount.\number\ssubcount.}
\sectcount=0

% macros for equation, figure and footnote numbering and naming
%
% count equations
%
\newcount\eqcount
\eqcount=0
\def\equno{\global\advance\eqcount by 1 \number\eqcount}
\def\equat#1{\count99=\eqcount \advance\count99 by #1 \number\count99}
%%%%%%%%%%%%%%%%%%%%%%%%%%%%%%%%%%%%%%%%%%%%%%%%%%%%%%%%%%%%%%%%%%%%%%%%%%%%%%
%                                                                            %
%    to name an equation, place command "\eqnam{\Poisson}" after equation,  %
%    and thereafter "equation(\Poisson)" will generate the proper equation   %
%    number.                                                                 %
%                                                                            %
%%%%%%%%%%%%%%%%%%%%%%%%%%%%%%%%%%%%%%%%%%%%%%%%%%%%%%%%%%%%%%%%%%%%%%%%%%%%%%
\def\eqnam#1{\xdef#1{\the\eqcount}}

%
% count figures
%
\newcount\figcount
\figcount=0
\def\nextfig{\global\advance\figcount by 1 Figure~\number\figcount}
\def\figno#1{\count99=\figcount \advance\count99 by #1 Figure~\number\count99}
\def\fignam#1{\xdef#1{\chaphead\the\figcount}}

%
% count footnotes
%
\newcount\notecount
\notecount=0
\def\note#1{\global\advance\notecount by 1
   \footnote{$^{\the\notecount}$}{\tenpoint #1\par}}

\def\refs{\parskip=0pt\sect{References}
\parindent=0pt\everypar{\hangindent 1cm}
\ifnum\linespacingstep=1 \tenpoint \fi

% ApJ Style
\def\AAp{{A\&A,} }
\def\AApS{{A\&AS,} }
\def\AJ{{AJ,} }
\def\AnnRev{{ARA\&A,} }
\def\ApJ{{ApJ,} }
\def\ApJL{{ApJ,} }
\def\ApJS{{ApJS,} }
\def\ApSS{{Astrophys. Sp. Sci.,} }
\def\BAN{{Bull. astr. Inst. Netherlands,} }
\def\Cambridge{(Cambridge: Cambridge University Press)}
\def\JCP{{J. Comput. Phys.\/} }
\def\MNRAS{{MNRAS,} }
\def\Kluwer{(Dordrecht: Kluwer)}
\def\Messenger{{ESO Messenger} }
\def\Nature{{Nature\/} }
\def\Obs{{Observatory,} }
\def\PASJ{{PASJ,} }
\def\PASP{{PASP,} }
\def\PhD{{PhD thesis,} }
\def\PhilTrans{{Phil. Trans. R. Soc. Lond. A,} }
\def\PhysFl{{Phys. Fluids,} }
\def\PhysRep{{Phys. Reports} }
\def\Reidel{(Dordrecht: Reidel)}
\def\RMP{{Rev. Mod. Phys.} }
\def\RPP{{Rep. Prog. Phys.} }
\def\SovAst{{Sov. Astron.} }
\def\SovAstL{{Sov. Astron. Letters} }
\def\Vistas{{Vistas Astron.} }

}

\def\ie{i.e.}
\def\eg{e.g.}

%% file: sect1.tex
\sect{Introduction}
The existence of global warps in galaxies has long resisted a convincing theoretical interpretation; Binney (1992) gives a recent summary both of the observations and theoretical work.  The principal difficulty, first noted by Kahn \& Woltjer (1959), arises from the radial variation of the unforced precession rates when the outer disk is tilted with respect to the inner parts which would cause a warp in a disk of test particles to wind up rapidly.  Lynden-Bell (1965) suggested that gravitational coupling in a massive disk might be able to overcome this tendency for warps to wind up but Hunter \& Toomre (1969, hereafter HT69), in a seminal paper, showed that the coupling forces for gradual bends are not strong enough to prevent rapid winding near the low surface density edge of a realistic disk where warps are observed.

To simplify their analysis, HT69 restricted their attention to a model of a disk galaxy in which the unperturbed orbits of all stars were precisely circular and coplanar.  In this approximation, they were able to establish two key results: that there were no {\it instabilities\/} that could give rise to warps ($m=1$, or to axisymmetric, $m=0$, bends either) and that the eigenfrequencies of almost all neutral bending modes become continuous in realistic disks with non-sharp edges.  This second result implies that any disturbance contains a broad range of frequencies and will therefore disperse over time; an impulsively induced warp in a cold disk would quickly degenerate into small-scale corrugations.

The second result was emphasized by Toomre (1983), who re-stated it in terms of the group velocity of bending waves.  Both results have been repeatedly cited (Binney \& Tremaine 1987, Sparke \& Casertano 1988, Binney 1992, Sparke 1995) as the reason warps cannot be explained as discrete oscillations of a galaxy.  However, they were established only in the approximation of a cold, razor thin disk and {\it neither\/} result carries over to more realistic disks in which the stars have random velocities!

The existence of bending instabilities in warm disks has been known for some time (Toomre 1966, Kulsrud, Mark \& Caruso 1971, Polyachenko 1977, Fridman \& Polyachenko 1984, Araki 1985, Malkov 1989, Sellwood \& Merritt 1994 [hereafter SM], Merritt \& Sellwood 1994 [hereafter MS]).  Axisymmetric instabilities are the most disruptive in warm disks, thickening the system in the direction normal to the plane (SM).  I show in \S4 that they are most vigorous in the inner parts of a realistic disk, where the in-plane random motions are expected to be highest, and can be suppressed in the outer parts by comparatively modest disk thickness.  While these instabilities could possibly play a role in the formation of bulges, an $m=1$ variant does not seem promising as an explanation for the origin of warps.

In \S5, I show that razor-thin disks with random motions in the plane in fact support a discrete spectrum of neutral modes.  Unfortunately, these neutral modes are generally strongly Landau damped in the still more realistic case of a finite thickness disk.  The damping occurs through coupling of the wave motion to the free vertical oscillations of the particles.  Thus the prospect of accounting for warps as discrete modes of oscillation of the disk remains bleak.

The only positive note sounded in this paper is the discovery, reported in \S6, that some fully self-gravitating, warm disks of finite thickness can support long-lived, large-amplitude, axisymmetric flapping oscillations.  While I am unable to offer a convincing explanation for this purely numerical result, if it is not an artifact it indicates that the physics of bending modes is not yet fully understood.  Even this may not deserve much attention, since Dubinski \& Kuijken (1995) and Nelson \& Tremaine (1995) find that bending oscillations are generally damped by live halos; the lifetimes of the discrete oscillations reported here will therefore depend on the damping rate by such a halo.  Thus these discrete oscillatory modes, intriguing though they may be to theoretical dynamicists, may be restricted to the highly idealized models which display them.

Because they occur in the absence of a halo, the bending modes discussed in this paper are quite different from the possible motions of a disk caused by a displacement or misalignment with its surrounding halo.  Oscillations of this type have been discussed for warps by Toomre (1983), Dekel \& Shlosman (1983), and Sparke \& Casertano (1988), while Sparke (1995) considers the axisymmetric equivalent, which she dubs a ``bowl mode''.\note{The survival of such modes has been called into question by the work of Dubinski \& Kuijken and of Nelson \& Tremaine.}

All the $N$-body simulations described in this paper are restricted to axial symmetry in order to permit much higher spatial resolution than is possible in fully 3-D work.  In \S2, I describe the $N$-body code devised for this purpose, and in \S4 I show that it is able to establish the bending stability boundary for axisymmetric global modes in situations where lower spatial resolution codes had failed (MS).

%% file: sect2.tex
\sect{Axisymmetric $N$-body Code}
In order to study axisymmetric instabilities with ample spatial resolution and without gravity softening, I have implemented an $N$-body code on a two-dimensional grid.  The grid points are uniformly spaced in both the radial and vertical directions, the vertical spacing being generally much less than the radial.  As usual, the grid is used to tabulate values of the gravitational field arising from masses distributed over the same set of points.  The gravitational field is smoothed on the scale of the grid spacing, of course, but with only two spatial dimensions it is easy to set the spacing to remain less than the scale on which the density varies.

The force components acting are convolutions of the mass on the grid points with the field of uniform surface density, thin, massive, ring-like elements having a finite radial extent, but zero vertical thickness.  Expressions for the gravitational attraction of such Saturn-ring like mass elements, without softening, are given in Appendix A and are pre-calculated and stored at the beginning of a run.  Grid smoothing is minimized by computing the radial and vertical acceleration components separately, rather than deriving them numerically from a tabulated gravitational potential.  The convolutions can be performed more efficiently in Fourier space, but only in the vertical direction.  The lowest resolution runs reported here used a grid having 100 nodes in the radial direction and 135 vertically, but these values were increased considerably whenever they were suspected of being inadequate.  I used a simple linear interpolation scheme between grid points.

In principle, just four coordinates need be advanced instead of the usual six, since the motion is constrained to be axisymmetric.  However, the resulting equations become difficult to treat numerically near $r=0$, as already noted by van Albada \& van Gorkom (1977), and I have therefore adopted standard leap-frog in Cartesian coordinates, with a time step typically 0.02 dynamical times.  

The initial positions and velocities of the particles (usually $100\,000$) are chosen as described in the next section.  Every particle is inserted twice with equal and opposite $z$-components of position and velocity (a quiet start) to ensure that the plane is initially flat and any subsequent departures are due to the growth of a bending instability.

The aspect ratio of the grid cells was chosen such that there were typically 5 points per Gaussian scale height, $z_0$, or about 50 points across the full thickness of the disk.  Reducing the vertical resolution caused the growth rates to decrease slowly, by about 5\% for a factor two reduction.  The radial resolution, while adequate for radially hot models, needed to be increased for cooler disks as the radial wavelength of the fastest growing mode becomes shorter.  Grid cells flatter than about 20:1 appear not to yield reliable results, probably because of the implicit large radial extent of the mass elements, and the number of radial grid points was again increased to avoid this problem.  Other tests showed that the results are insensitive to choices of the time step and particle number.

In many models, I applied a cylindrically symmetric radial force field, equal to the analytic expression in the mid-plane of an infinite zero-thickness disk, instead of computing the radial forces from the particle positions.  The advantages of doing this are two-fold: firstly, the same file of initial positions and velocities can be used for disks of differing thickness and secondly, it suppresses the axisymmetric Jeans instability which may saturate in the cooler disks before the bending mode can be measured accurately.  The growth of small-amplitude bends should not be affected by a fixed cylindrically symmetric force.  (A spherically symmetric force would affect the growth rate of the bend by providing an additional restoring force component towards the mid-plane.)

One of the principal results from each simulation is the growth rate of the dominant axisymmetric bending mode.  This is deduced from fitting an exponentially growing single function of radius to the mean $z$ displacements of the particles computed in a set of radial annuli at equal time intervals, as described in SM.

%% file: sect3.tex
\def\KT{{\tenrm KT}}
\sect{Kuz'min-Toomre models}
The models studied in this paper are a family of fully self-gravitating \KT\ disks of finite thickness.  These models are physically the same as those used in SM and MS and have the surface density distribution $$
\Sigma(R) = {M \over 2\pi a^2} \left( 1 + {R^2 \over a^2} \right)^{-3/2}, \eqno(\equno)
$$ where $M$ is a unit of mass and $a$ a length scale.  I truncate this infinite mass distribution by eliminating all particles with sufficient energy to reach radii larger than $6a$ in the majority of the models reported here, but increased the outer limit to $8a$ for the hottest disks in which the bending mode has a very large radial extent, and reduced it to $4a$ for the coolest models in which the outer parts are unaffected by bending instabilities.  A truncation in energy leads to smooth density fall-off anyway, but I further smooth the edge by tapering in angular momentum.

The particles are dispersed about the mid-plane in a Gaussian distribution with a scale height that is independent of radius for any one model.  A constant scale height was dictated by three considerations: first, it is simple with just one parameter; second, it is suggested by observations (\eg, van der Kruit \& Searle 1981) and third, a uniform thickness leads to a particularly simple change in the radial gravitational field in the mid-plane, simplifying the problem of generating a \DF.

A disk of finite thickness requires a three-integral \DF\ to generate equilibria with arbitrary axis ratios for the velocity ellipsoid.  Since no such functions are available, I adopt a two-integral \DF\ to set the initial velocity components parallel to the plane, and approximate the vertical equilibrium by solving the 1-D Jeans equation, as described in SM.  This procedure achieves a good equilibrium when the in-plane random motions are small, but hot models quickly adjust to a somewhat increasing thickness with radius.

Kalnajs (1976) works out detailed expressions for \DF s that reproduce the \KT\ disk density profile when the potential has the analytic expression for the infinite, zero-thickness disk.  Giving such models a finite thickness weakens the central attraction and therefore upsets the radial balance.  I have found that the gravitational field in the mid-plane of a uniformly thick \KT\ disk is well approximated by the simple expression (B1) with a single fitting parameter.  This is very convenient, since Kalnajs's prescription is easily adapted to other similar potentials; the expressions for the in-plane \DF\ in this approximate potential are given in Appendix B.  The \DF s have a single free parameter $\mk$ that determines the degree of random motion; $\mk = \infty$ for a cold disk (all particles on circular orbits) while $\mk=3$ for a disk that is non-rotating and isotropic in the plane.  The evolution in axisymmetric simulations is unaffected by the direction of the azimuthal velocity and no rule for retrograde particles is required.

In principle, it is necessary to determine a new \DF\ using the best fitting potential for each different initial $\mk$ and disk thickness.  For linear bending modes, however, it is sufficient to use the \DF s given by Kalnajs for the thin disk and to substitute a fixed cylindrical attraction of the required analytic form in place of the self-consistent radial attraction of the particles; I have taken advantage of this trick for most models reported in the next section where only the linear growth rate is of interest.

%% file: sect4.tex
\sectandsub{Bending instabilities}{Motivation}
Bending instabilities were first predicted in a local analysis by Toomre (1966).  In that terse and sequestered publication, he considered the bending modes of the Spitzer sheet, an infinite, constant surface density, non-rotating, slab of stars having Gaussian distributions of velocities both parallel to and perpendicular to the plane.  He showed first that when the scale of the bend was large compared with the slab thickness, the frequency, $\omega$, is related to the wavelength of the bend $\lambda = 2\pi/k$, through $$
\omega^2 = 2\pi G\Sigma |k| - k^2 \sigma_u^2, \eqno(\equno)
$$ where $\sigma_u^2$ is the in-plane dispersion of stellar velocities.  The horizontal motion is clearly destabilizing.  The ripple is oscillatory when $$
\lambda > \lambda_{\rm J} = {\sigma_u^2 \over G\Sigma}. \eqno(\equno)
$$ Smaller scale ripples would appear to be unstable, but finite thickness ($z_0 = \sigma_w^2 / G\Sigma$, \eg\ Binney \& Tremaine 1987, p282) clearly cannot be neglected in this regime.  Toomre concluded that the sheet would be stable on all scales only if $$
\sigma_w > 0.30 \sigma_u. \eqno(\equno)
$$  A more detailed re-analysis of the same problem by Araki (1985) yielded a value of 0.293 for the critical velocity dispersion ratio.  Another local stability analysis of a slab, having a less realistic vertical structure, was reported by Fridman \& Polyachenko (1984).  

As already noted, the global bending analysis in HT69 assumed cold disks, as did those of Sparke \& Casertano (1988) and Sparke (1995).  Polyachenko (1977), on the other hand, derived the complete spectrum of normal bending modes of the radially hot, but zero thickness, Kalnajs disks, finding that all, except the limiting case of exactly circular orbits, were highly unstable.  MS also report, in their semi-analytic study of bending modes, that the infinite, thin, \KT\ disk has a continuum of instabilities whenever random motion was added, and that both $m=0$ \& 1 instabilities were driven by random motion in the radial direction.  Both these global analyses were for razor-thin disks and were therefore unable to test how well the stability criterion (\equat0) held up in a global context.

A global analysis of bending modes in thickened disks would be a major challenge, requiring the derivation of a good \DF\ and the solution of the stability problem in one higher dimension.  SM and MS therefore used $N$-body simulations to study thickened disks with large degrees of random motion in the plane.  They found that the na\"\i ve Toomre-Araki stability criterion (\equat0) breaks down when the dominant bending mode has a radial wavelength comparable to, or greater than, the scale on which the surface density varies; \ie, where the local approximation is no longer valid.  They argued that the gravitational restoring forces at the disk center arising from the displaced outer disk are weaker than the local analysis assumes, requiring a greater degree of vertical pressure to stabilize these global modes.  Unfortunately, resolution limitations in their fully 3-D grid-based calculations prevented them from determining either the length scale on which the local criterion begins to fail, or how thick a disk needs to be for stability for a given degree of random motion in the plane.  In this section, therefore, I present results from much higher resolution numerical simulations restricted to axial symmetry which are designed to clear up both these outstanding issues.

\subsect{Numerical results}
Table 1 gives the growth rate of the most unstable bending mode, the only one which can be determined with any precision in this approach.  The first two columns give the \DF\ index $\mk$ and the quantity $r_*$ is that defined in Appendix B; when $r_*/a=1$, the radial forces were simply set from the analytic expression for the zero-thickness \KT\ disk, while $r_*>1$ indicates that the \DF\ used is computed for an excellent approximation to the field of the finite thickness disk.  The lengths $r_*$ and $z_0$ are in units of $a$, the velocities $\sigma_u$ and $\sigma_w$ are in units of $\sqrt{GM/a}$, while the frequency unit for the growth rate is $\sqrt{GM/a^3}$.

\nextfig\ shows the best-fit mode from four representative models.  The first three illustrate how quickly the radial wavelength of the dominant bending mode decreases with the magnitude of the radial motion in the plane.  It is therefore no surprise that the low resolution experiments reported by MS revealed instabilities in only the hottest of these models.  Both the shapes of the modes and their eigenfrequencies are insensitive to changes in the numerical parameters.

It seems inappropriate to describe the dominant instability as a ``bell'' mode (\cf\ Polyachenko 1977, MS) because the eigenfunction is clearly has a wave-like form, even for the fully pressure-supported disk.  In all but the hottest models, the lowest order bending mode has a very small amplitude away from the center; the distance to the first node shrinks as the radial velocity dispersion declines and the bend usually becomes undetectable well before the outer edge, unless the disk is very hot.  The outward decrease in amplitude in these thickened disks contrasts with the behavior seen in disks of zero thickness where the mode amplitude generally rises towards the outer edge, \eg, Figure 1 of MS.

I have not attempted to look for bending modes in still cooler disks because their short radial wavelengths would require yet higher radial resolution than used here.  Also the results for the cooler disks became more confusing because many modes grow and saturate with small limiting amplitudes at different times making it difficult to determine individual growth rates.

Thickening the disk always reduces the growth rate of the dominant mode and eventually appears to stabilize it -- or at least to reduce the growth rate so much that no instability can be detected.  The lowest order mode is not always the last to be stabilized, however.  Since the radial velocity dispersion in the plane generally drops less rapidly than the dispersion in the vertical direction, the local ratio $\sigma_w/\sigma_u$ decreases outwards and criterion (\equat0) could be violated in the outer disk.  (This does not necessarily imply instability; see below.)  The \KT/5 disk with a thickness of $0.08a$ is an example; in this model, $\sigma_w/\sigma_u \sim 0.5$ at the center, but drops below 0.3 by the edge.  \figno0(d) shows the dominant bending instability of this model; the evidence that it is not the lowest order mode is that its peak amplitude occurs away from the center.

\subsect{Global {\tenit versus\/} local stability boundary}
The growth rates of the lowest order mode from Table 1 are plotted in \nextfig, which can be compared directly with Figure 3(a) of MS obtained from a code having much lower spatial resolution.  The stability boundary given by (\equat0) is marked by the straight dashed line and the existence of strongly unstable models above this line confirms, with this higher quality code, the previous result that hot disks are much more unstable than predicted by a na\"\i ve application of Toomre-Araki local theory.  An approximate global bending stability boundary implied by these results is indicated by the dotted curve in \figno0, which asymptotes to the local stability line for cooler disks.  (The higher order instabilities are not plotted here because their stability is likely to depend more on the velocity dispersion ratio away from the center.)

Since the hottest ($\mk=3$) models are isotropic in the plane, or fully pressure supported, these results suggest that criterion (\equat0) fails significantly only in models in which radial pressure exceeds rotational support.  Merritt \& Sellwood (MS) pointed out that the breakdown of this criterion can be understood in terms of the reduced gravitational restoring force from distant parts of the disk when the surface density declines on a scale that is comparable to the radial wavelength of the mode.  The radial proportions of the lowest order bending modes in \figno{-1}\ make this suggestion appear still more plausible.

\def\VD{V_{\rm D}}
\def\kappaD{\omega_2}
\def\kappaH{\kappa_{\rm z,H}}
Toomre (private communication) has pointed out, however, that a purely local analysis omits an important term that was first identified by HT69 for disks with non-flat rotation curves.  HT69 found it useful to divide the vertical restoring force towards the mid-plane of their disk into two components which they denoted $F_1$ and $F_2$.  The $F_1$ part arises from the density corrugations themselves, while $F_2$ is the restoring force from the undisturbed matter.  At a small vertical distance above the plane, $F_2$ is clearly $z$ times the derivative of the vertical force just above the undisturbed plane.  At large radii, where the mass of the galaxy can be regarded as concentrated at $R=0$, we find $F_2 = -GMz/R^3$ and the restoring force increases with height.  Well inside the disk, on the other hand, $F_2$ will change sign where the vertical force decreases with height -- \eg, as it would in the extreme case in which all the mass is located immediately below the field point.  For matter distributed in a thin axisymmetric disk, applying Laplace's equation to the disk potential, $\nabla^2\Phi_{\rm disk}=0$, just above the thin disk allows $F_2$ to be related to the radial derivative of the disk potential (HT69) $$
F_2(R) = -z\left. {\partial^2 \over \partial z^2} \Phi_{\rm disk}(R,z) \right|_{z=0+} = \left. {z\over R} {\partial \over \partial R} \left( R {\partial \Phi_{\rm disk} \over \partial R}\right)\right|_{z=0} = {z\over R} {d \VD^2 \over dR}, \eqno(\equno)
$$ where $\VD(R)$ is the circular velocity in the mid-plane.  Thus, $F_2$ is zero where the rotation curve from the disk matter is flat, while it is positive (\ie, repulsive) where the rotation curve rises.

The final form in (\equat0) is correct only if the rotation curve arises {\it from the disk matter only\/}.  If a halo or other spherically distributed mass is also present, we must further divide $F_2$ into separate parts from the disk and halo.  The disk contribution to $F_2$ is still given by (5), as long as $\VD(R)$ is the contribution from the disk alone, but the halo matter will add yet another term $$
F_2(R) = {z\over R} {d \VD^2 \over dR} - z\left.{\partial^2 \over \partial z^2}\Phi_{\rm halo}(R,z)\right|_{z=o}. \eqno(\equno)
$$ It is convenient to write this as $F_2 = -z(\kappaD^2 + \kappaH^2)$.  Note that {\it neither\/} of these frequencies is $\kappaz \;(= \sqrt{\partial^2\Phi_{\rm total}/\partial z^2})$, the vertical epicycle frequency for small oscillations of a particle about the mid-plane.  The disk contribution to $\kappaz$ depends on the volume density in the disk mid-plane and is formally infinite in a razor-thin disk; $\kappaH$ is always the halo contribution to $\kappaz$, however.

Toomre's dispersion relation (\equat{-4}) was derived for an infinite sheet, for which $F_2=0$ (since $\nabla^2\Phi=0$ everywhere outside the sheet which is uniform in $x$ and $y$).  In order to apply it to realistic disks, we must therefore include the extra terms that come from the additional restoring forces, to obtain $$
\omega^2 = \kappaD^2 + \kappaH^2 + 2\pi G\Sigma |k| - k^2 \sigma_u^2, \eqno(\equno)
$$ with $\kappaH=0$ in this paper.  This refinement will clearly have the effect of increasing the critical wavelength $\lambda_{\rm J}$ for rising rotation curves, since $\kappaD^2<0$ in that case.  The stabilizing effect of finite thickness is not so easily adjusted for a different rotation curve slope, however, but we expect a greater thickness to be required for stability when $\lambda_{\rm J}$ exceeds that for a flat rotation curve.  The correction is large for hot disks, but the empirical stability curve asymptotes to the na\"\i ve Toomre-Araki line in very cool disks, because the typical $k$ increases as $\sigma_u$ drops and the importance of the constant correction $\kappaD^2$ dwindles.

In regions where the rotation curve declines, on the other hand, $F_2$ is stabilizing, and disks could still be locally stable with somewhat less vertical velocity dispersion than required by (\equat{-3}).

%% file: sect5.tex
\sect{Neutral modes in disks of zero thickness}
The lowest order instabilities found in \S4 are confined to inner parts of the disk unless it is very hot, and in these cases the instabilities become totally disruptive (SM).  Since bends are most commonly observed in the outer parts of galaxies where these instabilities have essentially zero amplitude in realistically cool disks, it makes sense to examine whether neutrally stable bending modes are more promising.  Neutral modes would need to be excited, of course, which could be achieved either externally, \eg, by the near passage of a companion galaxy, or perhaps internally through the non-linear evolution of instabilities in the inner disk.

As mentioned in the introduction, HT69 showed that the spectrum of neutral bending modes of a cold, thin disk is continuous, unless the outer edge is sharp, and that an impulsively applied bend will quickly wind up and also propagate towards the outer edge.  This widely quoted result is, however, true only in the extreme limit of all particles on circular orbits.  In this section, I show that razor-thin disks with random motion {\it do\/} support large-scale, discrete neutral modes, in addition to the unstable continuum first reported by MS.  Unfortunately, these discrete modes generally become strongly damped in the still more realistic case of a disk having sufficient thickness to suppress bending instabilities.

\subsect{Discrete spectrum in a disk of zero thickness}
The numerical scheme for determining the bending mode spectrum of a warm thin disk was described in MS.  Briefly, the integral equation for the normal bending modes is converted to matrix form by a procedure known as collocation.  In effect, the disk is divided into $n$ concentric ring elements which are each allowed to tip independently, but are coupled through the radial motions of the particles and by their mutual gravitational interactions.  The discrete mode spectrum of this finite set of coupled oscillators approximates that of the continuous disk.  Increasing $n$ (the number of collocation points or mass elements) naturally leads to a larger number of discrete modes; if the frequencies of these numerically determined modes become more densely packed every time the number of rings is increased, then the spectrum of the smooth disk is likely to be continuous.  On the other hand, the smooth disk can be deduced to have discrete modes if the eigenfrequencies of the low order modes retain similar spacing and the eigenfunctions do not change much as the mass elements are repeatedly subdivided.

The mode spectrum of a warm, but razor thin, infinite \KT/5 disk has both stable and unstable parts as shown in \nextfig.   Most modes are unstable and the frequencies of the instabilities change continuously as $n$ is increased.  The unstable spectrum is continuous, therefore, as reported by MS.  These instabilities are a consequence of zero thickness, since modest vertical thickness would quench them all (\S4). 

The neutrally stable spectrum is much less dense in this case.   As $n$ is increased, more stable modes appear at ever lower frequencies, but the highest few eigenfrequencies are nearly constant and the shapes of their eigenfunctions do not change.  Thus this infinite, warm, razor thin, disk supports a discrete spectrum of neutrally stable modes (in addition to the unstable continuum), a result that A.~Toomre (private communication) has kindly verified independently.

The bending mode spectra of infinite, thin disks having different degrees of radial motion show that cooler disks (a) have more densely packed neutral modes which (b) extend to higher frequencies.  Only in the limit of zero radial velocity dispersion does the neutral spectrum become continuous, however.  (No unstable modes exist in this limit either.)  Conversely, the stable spectrum becomes more sparse as random motion in the plane is increased and, as noted by MS, no stable modes remain in the limit of full pressure support (\DF\ is a function only of energy).

Truncating this infinite disk merely limits the number of stable modes; a disk of finite extent is unable to support arbitrarily low frequency bending waves.  I truncate the disk by eliminating all unperturbed orbits that extend beyond some finite radius, which causes the surface density to taper smoothly to zero at that radius (in a warm disk).  The neutral spectrum remains discrete but the small number of neutral modes present (independent of $n$, for $n$ sufficiently large) decreases as the truncation radius is reduced.  \nextfig\ shows the only three non-trivial, neutrally stable, bending modes in the thin \KT/5 disk when orbits extending beyond $R=6a$ are eliminated; the trivial displacement mode is also stable.  The number of discrete modes remains fixed and their frequencies and shapes scarcely change as $n$ is varied widely.

For completeness, I note that a large number unstable modes are still present when the disk is truncated, but the spectrum appears to be discrete since the modes of lowest growth rate are well separated and have frequencies independent of $n$.

\def\kcrit{k_{\rm crit}}
\def\sigmau{\sigma_{\rm u}}
\def\omegam{\omega_{\rm max}}

\subsect{Reason for the discrete spectrum of neutral bending modes in a thin disk}
The discontinuous change in the nature of the neutral spectrum when in-plane random motions are added to the disk is most easily understood in the WKBJ approximation.  The improved version (\equat0) of Toomre's (1966) original dispersion relation must also approximate the local relation between frequency and wavenumber for a bending mode in a disk galaxy when the wavelength is short compared to the scale on which $\Sigma$ and $\sigmau$ vary.  Since $\omega^2$ varies quadratically with $k$, a mode of (constant) frequency, $\omega$, has two possible positive wavenumbers (and two negative) provided that $\omega^2>\kappaD^2$ and less than the maximum possible value of the rhs.  The local dispersion relation therefore has both short- and long-wavelength branches, and the sign of the group velocity (given as usual by $\partial\omega/\partial k$) is opposite on the two branches and changes sign with $k$.

The value of the maximum will generally decrease outwards in the disk (see below), and therefore a bending wave travelling outwards at constant frequency will reach a point where the two branches merge.  At this point the wave will go over to the other branch which will carry it back towards the center; \ie, there is a turning point.  The WKBJ dispersion relation does not give any useful guide to the behavior at the center, but it seems reasonable to suppose that the wave will be reflected.  Since the sign of change $k$ is changed by the reflection off the center, the second half of the feed-back loop retraces the first half in the reverse direction.  Bending waves are therefore confined to a resonant cavity for which only certain discrete frequencies are permitted.

Note that in the exceptional case of no random motion in the radial direction ($\sigma_u=0$), the dispersion relations (\equat{-5}) and (\equat0) have no maxima; arbitrarily high frequencies correspond to real wavenumbers and no turning point exists -- an outgoing wave will continue outwards indefinitely.  Thus a cold disk is truly special and unrepresentative of the general behavior of a thin disk for two separate reasons: first, is has no instabilities and second, its neutral mode spectrum is continuous rather than discrete.

One logical gap in the above argument remains: it requires the maximum permitted bending frequency to diminish outwards.  To see that this will be true in all physically reasonable circumstances, we may rewrite equation (\equat{-5}) in dimensionless form using the variables $\zeta = k/\kcrit$ and $Q = \sigmau/\sigma_{\rm u,min}$ with $$
\kcrit = {2\pi \over \lambda_{\rm crit}} = {\kappa^2 \over 2\pi G\Sigma} \qquad \hbox{and} \qquad \sigma_{\rm u,min} \simeq \sqrt{2\over7}{2\pi G\Sigma \over \kappa} = {3.36G\Sigma \over \kappa}. \eqno(\equno)
$$ Ignoring the $\kappaD^2$ term, which is important only for small $k$ in a non-flat rotation curve, the local dispersion relation can be recast as $$
\omega^2 = \kappa^2 \left[ |\zeta| - \textstyle{2\over7}Q^2\zeta^2 \right], \eqno(\equno)
$$ from which it is readily determined that $\omegam = \sqrt{7\over8}\kappa/Q$.  Thus the condition that $\omegam$ decrease outwards requires only that $\kappa$ decrease faster than $Q$, and since $Q$ can reasonably be expected to be roughly constant while $\kappa$ varies inversely as the radius for a flat rotation curve, this condition is likely to be fulfilled in all galaxies.  (I am indebted to A Kalnajs for this last argument.)

Unfortunately, this feed-back loop for razor thin disks is most unlikely to survive in disks of finite thickness, as I discuss next.

%% file: sect6.tex
\sect{Neutral modes in a disk of finite thickness}
Although a thin, warm disk supports a discrete spectrum of neutral modes, its instabilities would quickly thicken it.  The more interesting question, therefore, is whether discrete neutral modes could be sustained in a realistically thick disk.  There is a strong reason to doubt that this would be possible, since the vertical oscillations of stars in a thickened disk are generally expected to damp a bending wave through resonant interactions (A Toomre, unpublished notes from 1983, Weinberg 1991, Binney 1992).\note{Such resonances are absent in a zero thickness sheet, since the frequency of vertical oscillation is formally infinite.}

Once again, generalization of the global analysis of \S5 to disks of finite thickness would be a major challenge and would still yield only linear modes.  It is both easier and more interesting, therefore, to look for large-amplitude, long-lived waves in $N$-body simulations.

\subsect{Flapping modes}
In many simulations of unstable disks (\eg, SM), the models settle to a steady, rather than oscillating, equilibrium after the instabilities have run their course.  This result is precisely what would be expected if disturbances excited by saturation of the dominant instability were strongly damped.

On closer examination, however, there were a few cases in which the disk continued to quiver in a long-lived, large-scale oscillation, albeit at rather low amplitude.  After the disk had thickened and transients excited by the instability had decayed, the oscillation maintained a steady frequency and  showed little tendency to decay.  The amplitudes of these quiverings were not large ($\sim 0.035a$ peak to peak), but were well resolved by the grid.

Disks can oscillate with much greater amplitudes, however, as illustrated in \nextfig.  The \KT/5 model in this experiment was just thick enough ($z_0 = 0.1a$) to be stable, but was perturbed by the initial vertical displacement applied to each particle $$
{z_{\rm disp} \over a} = A \left[ {1 \over 2} - \left( 1 + {R^2 \over 4a^2} \right)^{3/2}\right], \eqno(\equno)
$$ with the amplitude $A=1/3$ in this case.  (This form was found from experience to minimize initial transients; the fundamental mode has a node near $R=2a$.)  \figno0\ illustrates just one complete oscillation of this perturbed model.  The large-amplitude bending oscillation persists right from the outset and decays quite slowly, as shown in \nextfig(a).  The best fit mode to these data has the eigenfrequency $0.080$; the oscillation frequency is well determined but the decay rate, though clearly very low is quite uncertain, even when the model was run for quite large numbers of oscillation periods.  The shape of this eigenmode is illustrated in \figno0(b); it does not have a constant phase at all radii in the sense that the oscillation in the outer parts lags that of the bulk of the inner disk, as may also be seen in \figno{-1}.

The slow decay of this oscillation was a real surprise, and at first I regarded it with suspicion.  I have subsequently checked that it did not change in any significant way when the numerical parameters (grid size, and number of particles, \etc) were varied.  As a more independent check, I tried running the same physical model on the 3-D Cartesian grid used by SM with the highest practicable spatial resolution.  In order to suppress non-axisymmetric instabilities, I computed the motion of each particle subject to acceleration components in the radial and vertical directions only.  The grid size in this case is $257^2 \times 65$ and the disk Gaussian scale height $0.06a = 1.2$ mesh spaces.  Once again the plane flapped coherently with a mode shape that was remarkably close to that in the physically identical model on the other grid with a frequency approximately 7\% percent lower.  That the two different grids supported the same flapping wave with frequencies that are even in reasonable quantitative agreement is reassuring.

\subsect{Vertical resonances}
The possible existence of long-lived oscillations in stellar systems has been discussed by several authors.  Sridhar (1989), Mathur (1990), Weinberg (1991, 1994) and Louis (1992) all discuss oscillations in one-dimensional or spherical stellar systems while Miller \& Smith (1994) present a spherical $N$-body model that oscillates for a long period.  Toomre (unpublished) and Weinberg (1991) also find that plane waves in an infinite sheet are generally strongly damped, although damping weakens for long wavelengths.  The present model is the first example of a global oscillation in two-dimensions of which I'm aware.  As emphasized by Binney (1992), a long-lived oscillation would be expected to occur at a frequency with which very few particles can resonate.

The radial variation of the vertical oscillation frequency of free, small amplitude vibrations of particles about the mid-plane, $\kappaz$, is plotted in \nextfig\ for some of the present models.  These curves were determined from the initial, unperturbed density distributions in the simulations.  The family of four full-drawn curves are drawn for four models of differing thicknesses which all supported long-lived vibrations similar to that shown in \figno{-1}\ but at the separate frequencies given in Table 2.  These models had the identical \KT/5 \DF\ for the in-plane velocities and were all truncated at $R=6a$ and given the initial kick of equation (\equat0) with $A=1/3$.\note{The thinner two of these models are mildly unstable when not perturbed (Table 1), but I could not find the instabilities in these perturbed models, probably because transient heating by the initial perturbation pushed them closer to the stability boundary.}

As the disk is made thicker, the free vertical frequencies of the particles at any radius decline, as also does the frequency of this large-scale flapping mode.  This plot would seem to suggest that no resonance can exist between the particles and the flapping mode in any of these cases, since the flapping frequency is always well below the minimum $\kappaz$ inside the disk edge.  It should be noted, however, that particles that make larger vertical excursions have a vertical frequency that is lower than $\kappaz$, and the possibility of resonance is not excluded by this plot.  Moreover, the changing phase of the oscillation towards the outer edge is a strong indication that a resonance is being approached.  Nevertheless, the fraction of particles in the disk that are in resonance with the flapping mode is clearly very small which would appear to account for the weak damping.

Since possible weak damping depends critically on an adequate density of particles in phase space, I have checked that a ten-fold increase in particle number (to 1 million particles) does not change the result.  The frequencies in the two physically identical runs differed by less than a percent while the amplitude at late times was actually higher by a few percent in the large $N$ case.

One possible mechanism for the oscillation could be that waves are reflecting off the outer edge; HT69 had shown that a cold, thin disk could be made to oscillate in discrete modes if the outer edge were sharp enough to reflect out-going WKBJ waves in a finite time (see also Toomre 1983).  The edge of the disk used in the present experiments was quite fluffy in their sense, but probably what is more important is the density gradient normalized by the scale of the wave; since the mode has such a large scale, it is possible that the edge is sharp enough to reflect.  I therefore tried changing the outer limiting radius of the disk, keeping all else, including the form of the initial perturbation fixed.  Both increasing the limiting radius to $8a$ and decreasing it to $4a$ made surprisingly little difference to the flapping mode.   The oscillation frequency did not change within the errors of measurement and the approximate position of the ``node'' moved inwards somewhat in the more abruptly truncated disk only, while the out of phase region was larger in the more extended disk.  These two results strongly suggest that reflection off the outer edge is unimportant for this mode.

The {\it damping\/} rate, on the other hand, was substantially increased by extending the disk to larger radii; while the vibration frequency was hardly changed in this case, the damping rate was quite clearly non-zero and the amplitude of the oscillation almost halved every period.  (The oscillation in the still more abruptly truncated disk did not decay significantly.)  The two dashed curves in \figno0\ show $\kappaz$ for these two models.  The number of particles that can resonate with a wave of fixed frequency is clearly higher in the more extended disk, where the out of phase region of the mode was more extensive and the damping rate higher.

The dotted curve shown in \figno0\ is for the same disk that was still thicker, and as indicated in Table 2, the initial perturbation in this case was damped very quickly with little sign of oscillation.  Resonance damping could have been avoided in this model at a still lower pattern speed, yet the model seemed unable to support a mode of still lower frequency.  I speculate on the reason for this in \S6.4.

\subsect{A non-linear effect}
The change in behavior caused by changes in the amplitude of the initial kick given to the system suggests that these could be non-linear oscillations.  I conducted a series of experiments with the same equilibrium model, the \KT/5 disk truncated at $R=6a$ with $z_0=0.1$, but in which I varied the amplitude, $A$, of the initial kick.  The ``final'' amplitude is always lower than the applied perturbation, and decreases faster than linearly with $A$, \ie\ some damping occurs initially.  Since the wave decayed very slowly, if at all, over the period $200 \leq t \leq 300$, I measured the amplitude of the best fit wave with a purely real frequency over this period.  Table 2 reports the frequency of this fitted wave, and its amplitude, $B$, reckoned as the mean of the absolute amplitudes over the radial range $0 \leq r/a \leq 4$.  For very small initial kicks ($A < 1/15$) the oscillation damps so much in the early evolution that it is difficult to detect at all at later times.

This behavior appears to indicate that some resonant particles are present, but that their ability to damp large perturbations is limited -- \ie, the resonance saturates.  This could be either because all the resonant particles are scattered out the resonant range, or that they become trapped by the large-amplitude wave.

\subsect{Radial velocity dispersion}
Reducing the magnitude of the radial velocities, while keeping the truncation radius constant, quite clearly led to much stronger damping.  I could find little evidence for true oscillations in \KT/12 disks even when I reduced the thickness considerably, and oscillations in a \KT/8 disk lost some 80--90\% of their amplitude in each cycle.  Thus a rather small reduction in the sizes of typical epicycles leads to a substantial reduction in the ability of this disk to oscillate.

A possible explanation of this change of behavior is that the only radially very hot disks can support oscillations at a low enough frequency to avoid strong vertical resonances.  The thin-disk dispersion relation (7) may give some indication of the frequency at which a thickened disk can oscillate over the range $a \ll 1/k \ll z_0$ -- \ie, for wavelengths much longer than the disk thickness and shorter than the radial scale on which the surface density changes.  The oscillation shown in \figno{-1}(b) could lie in this narrow range of scales.  Low frequency waves of such small $k$ probably require $\sigmau$ to be large in order to make the negative term significant.  Thus, only hot disks seem able to support low enough frequencies to avoid vertical resonances.

If this is correct, it may also account for the abrupt change in the ability of the \KT/5 disk to oscillate when the thickness was increased to $z_0=0.15a$.  To avoid damping, the frequency must be low enough to avoid vertical resonances.  This requirement becomes still more severe as the disk thickness is increased, and may cross a value below which no frequency can be supported at any $k$.

\subsect{Discussion}
The long-lived bending oscillations described in this section present a considerable theoretical challenge.  The evidence just presented indicates that they largely avoid the wave-particle interactions where Landau-type damping could have occurred at vertical resonances.  But I have not been able to identify a mechanism for the discrete mode.

The mode bears little resemblance to any of the linear modes of the razor-thin, finite disk illustrated in \figno{-3}.  Moreover, it cannot be expected to operate through a similar long- and short-wave feed-back cycle, since simple short waves clearly cannot be sustained in a disk that is significantly thick, even where Landau damping is avoided.  The experiments which showed that the mode frequency was unaffected by the position of the outer edge strongly suggest that a reflection off the edge is unimportant.

I have so far found long-lived flapping modes in only the \KT/5 models.  I have tried a few perturbed \KT/8 models which have smaller random velocities in the plane and find that they damp rapidly; perturbations in yet cooler (\KT/12) and thinner disks are super-critically damped -- \ie, they decay without oscillating at all.  I tried reducing the truncation radius in a \KT/8 disk and found that the damping rate was reduced somewhat.  Much more extensive experimentation (different disk thicknesses, truncation radii and initial perturbation amplitudes at least) is required, however, before I could rule out the possibility that these disks could be made to flap.

%% file: sect7.tex
\sect{Summary}
Most of the work reported in this paper further reinforces the conclusion of HT69 that it is hard to account for a gentle warp in an isolated disk galaxy as a discrete mode of the system.  Nevertheless, it hopefully clarifies some aspects of the effect of random motion on the bending behavior of disks.  In particular, I have shown that the frequently presented argument that disks lack discrete bending modes because the spectrum of neutral modes is continuous, is incorrect.

That disks with random motion have to be finitely thick to be stable to bending modes has been known since Toomre's (1966) pioneering local analysis.  However, it was not widely known that razor-thin disks with random motion also support a discrete spectrum of neutral modes.  Unfortunately, this little-known fact is of minimal significance, since the same vertical random motions that are required to stabilize the disk also cause nearly all imposed bends to decay through wave-particle interactions in a manner analogous to Landau damping.  Thus a realistic disk with both in-plane and vertical random motion would appear to have the same two properties that HT69 established for the idealized cold, razor thin case, viz: when a disk is thick enough to inhibit both $m=0$ and $m=1$ bending instabilities, discrete bending modes are not {\it expected\/} either.  The introduction of random motions has considerably complicated the problem, therefore, without changing the conclusion that bends are very hard to sustain!

In contrast to these rather negative general conclusions, I also report the discovery of a few exceptional cases of warm stable (in axial symmetry only) disks that were able to support long-lived flapping oscillations.  I am unable to offer a mechanism for these discrete modes, but they do appear to be real; tests with changes to the numerical parameters have revealed no indication that the mode could be a numerical artifact (though this does not constitute a proof).  Fascinating as this mode, if real, may be to theoretical dynamicists, its existence in only hot, fully self-gravitating disks would appear to render it unimportant for real galaxies.  Moreover, if a similar discrete bending oscillation could be found in a disk embedded in a rigid halo, it would probably be damped by a live halo (Dubinski \& Kuijken 1995).

The conclusion from this paper is that any possible window which would allow warps to be explained as discrete bending modes of the disk has been narrowed still further, but cannot be totally closed until the discrete oscillation reported here is understood.  Other suggested solutions, such as off-axis infall (Binney 1992) or of excitation by a halo (Nelson \& Tremaine 1996) are unaffected by the present work and therefore appear more attractive. 

It should be borne in mind, that this study is very restricted: disturbance forces in all the experiments were restricted to axial symmetry, all the disks had the same surface density and were fully self-gravitating -- \ie, none had a halo.  The results of \S4 mapped out, for the \KT\ disk, the approximate bending stability boundary for global instabilities; they confirm that the Toomre-Araki criterion derived in the local approximation is adequate when the disk is predominantly supported by rotation, and fails only when random motion plays a dynamically important role.  Non-axisymmetric instabilities, which are precluded from these experiments, are unlikely to give rise to a worse violation of the local criterion because the azimuthal dispersion is lower than the radial and surface density varies only in the radial direction.  The applicability of the local criterion to the rotationally supported outer parts of disk galaxies has been tested for a more realistic mass distribution (Sellwood 1996), and confirms Toomre's (1966) conclusion that the disk of the Milky Way is sufficiently thick to be well clear of the bending stability boundary.

%% file: appendA.tex
\sect{Appendix A}
The basic axisymmetric mass element is a uniform hoop.  The gravitational potential at an arbitrary field point, $(R,z)$, of a thin wire ring of radius $R^\prime$ and mass $\delta M$ lying in the $z=0$ plane centered on $R=0$ is $$
\delta\Phi(R,z) = -{G \delta M \over \pi \sqrt{RR^\prime}}  \alpha \, K(\alpha), \eqno(\rm A1a)
$$ where $K$ is a complete elliptic integral of the first kind and $$
\alpha^2 = {4RR^\prime \over (R+R^\prime)^2 + z^2}; \eqno(\rm A1b)
$$ (see \eg\ Binney \& Tremaine 1987, pp 72--73).  The radial acceleration is $$
\delta a_R(R,z) = {G\delta M \over \pi \sqrt{RR^\prime}} \left[ {E(\alpha) \over 1- \alpha^2} {\partial  \alpha \over \partial R} - { \alpha \, K(\alpha) \over 2 R} \right] \eqno(\rm A2a)
$$ where $E$ is a complete elliptic integral of the second kind and $$
{\partial  \alpha \over \partial R} = { \alpha^3 \over 8R} \left[ {R^\prime \over R} - {R \over R^\prime} + {z^2 \over RR^\prime} \right]. \eqno(\rm A2b)
$$ The vertical acceleration is $$
\delta a_z(R,z) = {G\delta M \over \pi \sqrt{RR^\prime}} \; {E(\alpha) \over 1- \alpha^2} \; {\partial  \alpha \over \partial z} \eqno(\rm A3a)
$$ with $$
{\partial  \alpha \over \partial z} = -{ \alpha^3 z \over 4RR^\prime}. \eqno(\rm A3b)
$$ These expressions are well defined everywhere except for the radial force and potential on the ring of mass itself, where $\alpha=1$.  

Since the central attraction of a massive ring on itself cannot be neglected, I give the rings a finite radial extent, imagining them to have uniform surface density (\ie, $\delta M = 2\pi\Sigma R\delta R$) and to extend mid-way to the next grid point on either side, while having zero vertical thickness.  The gravitational field of such a Saturn ring-like element of mass $M$ is $$ \eqalign{
{{\cal F}(R,z) \over M} & = \int_{R_1}^{R_2} d{\cal F} \Bigg/ \int_{R_1}^{R_2} dM \cr
& ={2 \over (R_2^2 - R_1^2)} \int_{R_1}^{R_2} {\delta {\cal F} \over \delta M} R^\prime dR^\prime, \cr} \eqno(\rm A4)
$$ with ${\cal F} = \Phi$, $a_R$ or $a_z$.  These expressions are finite everywhere, but the Cauchy principal value of the integral is required for the potential and radial attraction when $z=0$ and $R_1<R<R_2$.

In order to ensure that the vertical forces between any pair of mass elements on the grid are equal and opposite, I evaluate the integral for each pair of grid points just once and use the negative of the first result in place of the other.  (The difference between the two values arises from not evaluating the mass-weighted average of the force over the extended mass at the field point; the considerable extra expense of making this tiny adjustment seems unjustified.)

%% file: appendB.tex
\sect{Appendix B}
The gravitational potential in the mid-plane of a finite thickness, \KT\ disk of mass $M$ and length scale $a$ is well approximated by the expression $$
\Phi(R) = GM (R^2 + a^{\prime2})^{-1/2}, \eqno(\rm B1)
$$ where $a^\prime$ is a fitting parameter that is somewhat larger than $a$ and tends to $a$ in the limit of zero thickness.

Two-integral \DF s can easily be derived for this potential using the prescription given by Kalnajs (1976).  He introduces the dimensionless variables $$
x = - {h\sqrt{-2E} \over r_*\Phi(0)} \qquad \hbox{and} \qquad e = {E \over \Phi(0)}, \eqno(\rm B2)
$$ with $h$ and $E$ being the specific angular momentum and energy of a star, and $$
r_* \equiv \lim_{R\to\infty} {R\Phi(R) \over \Phi(0)}. \eqno(\rm B3)
$$ For the potential (B1), $r_*=a^\prime$.

In these variables, the phase space density takes the form $$
f(E,h) = e^{\mk-1}g(x), \eqno(\rm B4)
$$ where $$
g(x) = -{(2/x)^{\mk} \over 2\pi\Phi(0)} \left\{P_{\mk-1}(1)x {\partial \over \partial x}T(x) - {\mk (\mk-1) \over 2}T(x) + \int_0^1 P^{\prime\prime}_{\mk-1}(\eta) T(\eta x) d\eta \right\}, \eqno(\rm B5)
$$ with $P_n(\theta)$ being the usual Legendre polynomial and the double prime denoting the second derivative.  The function $T(y)$ is most easily evaluated in terms of the desired disk surface density $$
T(y) = \left( {R \over r_*} \right)^{\mk} \Sigma(R), \eqno(\rm B6a)
$$ where the radius is defined implicitly through $$
y = {R\Phi(R) \over r_*\Phi(0)}. \eqno(\rm B6b)
$$ For the \KT\ surface density in the potential (B1) $$
g(0) = -{\mk \over 2\pi^2\Phi(0)}.  \eqno(\rm B7)
$$ Since the expression (B5) is cumbersome and time-consuming to evaluate, I find it advantageous to obtain approximate values of $g(x)$ by interpolation in a pre-calculated table of 1000 logarithmic values spanning the range $0\leq x \leq x_{\rm max}$.  The value of $x_{\rm max}<1$ for a finite disk.

Kalnajs (1976) proves that expression (B5) is always positive when $\mk \geq 3$ and $r_* = a$.  In the softened potentials used here, $r_* >a$ and $g(x)$ can become negative for $\mk \gtsim 3$ as $x \rightarrow 1$.  Values of $x$ close to unity are avoided by truncating the disk, and the \DF\ has been verified to remain positive everywhere it is needed in this paper.

%% file: table1.tex
\sect{Table 1}

$$\vbox{\halign{
\indent # \hfil & \quad # \hfil & \quad # \hfil & \quad # \hfil & \quad # \hfil & \quad # \hfil \cr
\multispan6 {\hfil{\authorfont Unstable Models}\hfil} \cr
\noalign{\smallskip}
\multispan6 \hrulefill \cr
\noalign{\smallskip}
$m_K$ & $r_*$  &  $\sigma_u$ & $z_0$ & $\sigma_w$ & growth rate \cr % Run #
\multispan6 \hrulefill \cr
3 & 1.0      & 0.500 & 0.05  & 0.167 & $0.391\pm0.003$ \cr % 713
3 & 1.078    & 0.481 & 0.1   & 0.222 & $0.185\pm0.002$ \cr % 665
3 & 1.146    & 0.465 & 0.2   & 0.277 & $0.079\pm0.001$ \cr % 670
\noalign{\smallskip}
4 & 1.0      & 0.447 & 0.025 & 0.125 & $0.585\pm0.005$ \cr % 711
4 & 1.043    & 0.438 & 0.05  & 0.172 & $0.240\pm0.001$ \cr % 682
4 & 1.078    & 0.431 & 0.1   & 0.223 & $0.084\pm0.000$ \cr % 683
\noalign{\smallskip}
5 & 1.0      & 0.408 & 0.025 & 0.125 & $0.487\pm0.001$ \cr % 710
5 & 1.0      & 0.408 & 0.05  & 0.166 & $0.212\pm0.002$ \cr % 705
5 & 1.042    & 0.400 & 0.06  & 0.179 & $0.091\pm0.001$ \cr % 953
5 & 1.042    & 0.400 & 0.08  & 0.203 & $0.03~\pm0.01~$ \cr % 955
5 & 1.078    & 0.393 & 0.1   & 0.223 & undetectable    \cr % 686
\noalign{\smallskip}
6 & 1.0      & 0.378 & 0.025 & 0.125 & $0.387\pm0.003$ \cr % 708
6 & 1.0      & 0.378 & 0.05  & 0.168 & $0.116\pm0.003$ \cr % 702
\noalign{\smallskip}
8 & 1.0      & 0.333 & 0.0125& 0.093 & $0.495\pm0.003$ \cr % 714
8 & 1.0      & 0.333 & 0.025 & 0.125 & $0.209\pm0.001$ \cr % 700
8 & 1.0      & 0.333 & 0.05  & 0.168 & undetectable \cr    % 701
\noalign{\smallskip}
10 & 1.035   & 0.296 & 0.0125& 0.093 & $0.23 \pm0.02$ \cr % 690
10 & 1.0     & 0.301 & 0.025 & 0.125 & $0.04 \pm0.01$ \cr % 698
10 & 1.045   & 0.295 & 0.05  & 0.172 & undetectable \cr   % 673
\noalign{\smallskip}
15 & 1.0     & 0.250 & 0.0125& 0.093 & $0.099\pm0.002$ \cr % 712
\noalign{\smallskip}
30 & 1.0     & 0.180 & 0.004 & 0.053 & $0.17 \pm0.05$ \cr % 745
30 & 1.0     & 0.180 & 0.005 & 0.060 & undetectable \cr % 744
\noalign{\medskip}
}\hrule}$$

%% file: table2.tex
\sect{Table 2}

$$\vbox{\halign{
\indent # \hfil & \quad # \hfil & \quad # \hfil \cr
\multispan3{\hfil{\authorfont Varying Disk Thickness}\hfil} \cr
\noalign{\medskip}
\multispan3 \hrulefill \cr
\noalign{\medskip}
$z_0$ & $\sigma_w$ & eigenfrequency \cr % Run #
\multispan3 \hrulefill \cr
\noalign{\smallskip}
0.06  & 0.182 & $0.114\pm0.003 - (0.001\pm0.002)i$ \cr % 950
0.08  & 0.203 & $0.098\pm0.004 - (0.002\pm0.001)i$ \cr % 956
0.10  & 0.221 & $0.080\pm0.003 - (0.000\pm0.004)i$ \cr % 958
0.12  & 0.235 & $0.063\pm0.002 - (0.001\pm0.004)i$ \cr % 959
0.15  & 0.253 & damped    \cr % 957
\noalign{\medskip}
}\hrule}$$

%% file: table3.tex
\sect{Table 3}

$$\vbox{\halign{
\indent # \hfil & \quad # \hfil & \quad # \hfil \cr
\multispan3 {\hfil{\authorfont Varying Initial Amplitude}\hfil} \cr
\noalign{\medskip}
\multispan3 \hrulefill \cr
$30 \times A$ & $B$ & frequency \cr % Run #
\multispan3 \hrulefill \cr
\noalign{\smallskip}
10 & 0.0241  & 0.077 \cr % 1015
6  & 0.0159  & 0.086 \cr % 1013
4  & 0.0073  & 0.086 \cr % 1014
2  & 0.0010  & 0.102 \cr % 1012
1  & weak  \cr % 1016
\noalign{\medskip}
}\hrule}$$

%% file: refs.tex
{\refs

Araki, S., 1985, \PhD MIT

Binney, J., 1992, \AnnRev {\bf 30}, 51

Binney, J. \& Tremaine, S., 1987, {\it Galactic Dynamics}, (Princeton: Princeton University Press)

Dekel, A. \& Shlosman, I., 1983, In {\it Internal Kinematics and Dynamics of Galaxies} IAU Symposium {\bf 100} ed E. Athanassoula \Reidel\ p~187

Dubinski, J. \& Kuijken, K., 1995, \ApJ {\bf 442}, 492

Fridman, A. M. \& Polyachenko, V. L., 1984, {\it Physics of Gravitating Systems}, Springer-Verlag, New York

Hunter, C. \& Toomre, A., 1969, \ApJ {\bf 155}, 747 (HT69)

Kahn, F. D. \& Woltjer, L., 1959, \ApJ {\bf 130}, 705

Kalnajs, A. J., 1976, \ApJ {\bf 205}, 751

Kulsrud, R. M., Mark, J. W-K. \& Caruso, A., 1971. \ApSS {\bf 14}, 52

Louis, P. D., 1992, \MNRAS {\bf 258}, 552

Lynden-Bell, D., 1965, \MNRAS {\bf 129}, 299

Malkov, E. A. 1989, {\it Astron. Zh.}, {\bf 66}, 1189; English translation: \SovAst {\bf 33}, 614

Mathur, S. D. 1990, \MNRAS {\bf 243}, 529

Merritt, D. \& Sellwood, J. A., 1994, \ApJ {\bf 425} 551 (MS)

Miller, R. H. \& Smith, B. F., 1994, {\it Cel. Mech. Dyn. Astron.}, {\bf 59}, 161

Nelson, R. W. \& Tremaine, S., 1995, \MNRAS {\bf 275}, 897

Nelson, R. W. \& Tremaine, S., 1996, In {\it Gravitational Dynamics}, ed. O. Lahav, E. Terlevich \& R. Terlevich (Cambridge: Cambridge University press), (to appear)

Polyachenko, V. L., 1977, {\it Pis'ma Astron. Zh.}, {\bf 3}, 99; English translation: \SovAstL {\bf 3}, 51

Sellwood, J. A., 1996, In {\it Unsolved Problems of the Milky Way\/} (IAU Symposium {\bf 169}), Ed.\ L Blitz (Dordrecht: Kluwer) (to appear)

Sellwood, J. A. \& Merritt, D., 1994, \ApJ {\bf 425} 530 (SM)

Sparke, L. S., 1995, \ApJ {\bf 439} 42

Sparke, L. S. \& Casertano, S., 1988, \MNRAS {\bf 234}, 873

Sridhar, S., 1989, \MNRAS {\bf 238}, 1159

Toomre, A., 1966, {\it Geophysical Fluid Dynamics}, notes on the 1966 Summer Study Program at the Woods Hole Oceanographic Institution, ref. no. 66-46, p~111

Toomre, A., 1983, In {\it Internal Kinematics and Dynamics of Galaxies}, IAU Symposium {\bf 100}, ed E. Athanassoula \Reidel\ p~177

van Albada, T. S. \& van Gorkom, J. H., 1977, \AAp {\bf 54}, 121

van der Kruit, P. C. \& Searle, L., 1981, \AAp {\bf 95}, 105

Vandervoort, P. O., 1970, \ApJ {\bf 161}, 87

Weinberg, M. D. 1991, \ApJ {\bf 373}, 391

Weinberg, M. D. 1994, \ApJ {\bf 421}, 481

\par}

%% file: captions.tex
\sect{Figure Captions}

{\parindent = 0pt

\bigskip
\item{1} Radial profiles of the best fitting, small-amplitude, axisymmetric bending modes from disks with different properties.  (a) A radially hot model (\KT/4) with a large initial thickness ($z_0=0.1a$), (b) a $Q\sim1$ (\KT/8) and thinner ($z_0=0.025a$) case, (c) a very cool (\KT/30) model with $z_0=0.004a$ and (d) a higher order mode in a \KT/5 model with $z_0=0.08a$, which is sufficient to suppress the dominant mode at the center.

\item{2} Growth rates of the dominant bending mode from many models plotted as functions of the radial and vertical velocity dispersion at the center of the disk.  The sizes of the circles vary in proportion to the estimated growth rate (the scale is indicated top left) while crosses indicate that no instability was detected in the simulation.   The existence of non-zero growth rates above the dashed line shows that the locally derived Toomre-Araki stability boundary is strongly violated in some of these global simulations.  The dotted curve shows an approximation to the actual stability boundary, as judged from these results.

\item{3} Eigenfrequencies of axisymmetric modes of the infinite \KT/5 disk as $n$ (the number of collocation points) is varied.  Unstable modes are marked with pluses, stable modes as circles.  The unstable spectrum is continuous, since no well isolated, constant frequency emerges as $n$ is increased.  On the other hand, the few highest frequencies in the stable spectrum do not change with $n$, indicating they are discrete modes.  The behavior of the additional modes which accumulate at ever lower frequencies as $n$ increases seems to indicate that the entire neutral spectrum of the smooth disk is discrete.

\item{4} Eigenfunctions of the three neutrally stable modes in the \KT/5 which is cut off at $r=6a$.  The radial shapes of these modes do not change significantly once the number of elements is greater than 150, but the frequencies change in the third decimal.

\item{5} A small part of the evolution of the perturbed model discussed in \S6.1.  One particle in 20 in this axisymmetric simulation is plotted in meridional projection and the time range covers almost one complete oscillation period.

\item{6} (a) The time dependence of the mean displacement of the mid-plane in the radial range $2a\leq R \leq 4a$ of the perturbed model discussed in \S6.1.  (b) The best fit eigenfunction to the mean displacement of the disk over the period $40\leq t \leq 300$; the solid line shows the real part, the dashed line shows the imaginary part and the two dotted lines show plus and minus the amplitude of the mode.

\item{7} The radial variation of the frequency of small amplitude oscillations in the undisturbed mid-plane of most of the models reported in \S6.2.   The thickness of the disks with the dotted curve and four full-drawn curves were (from bottom to top) $z_0=0.15$, 0.12, 0.10, 0.08 and 0.06, and the two dashed curves relate to models with $z_0=0.6$ also, but truncated at $R=4a$ or $R=8a$.  All these models flapped, with the exception of that marked with the dotted curve in which the disturbance decayed without oscillating.

}
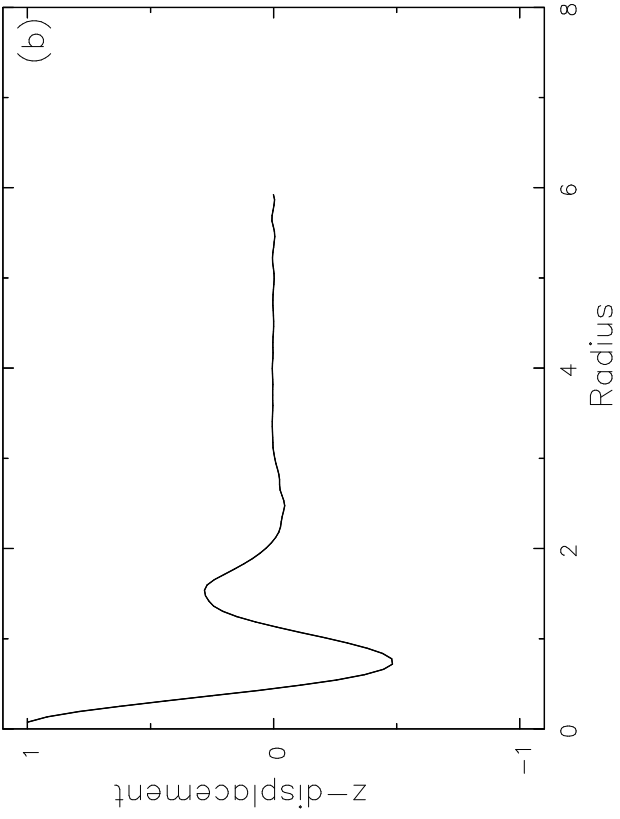
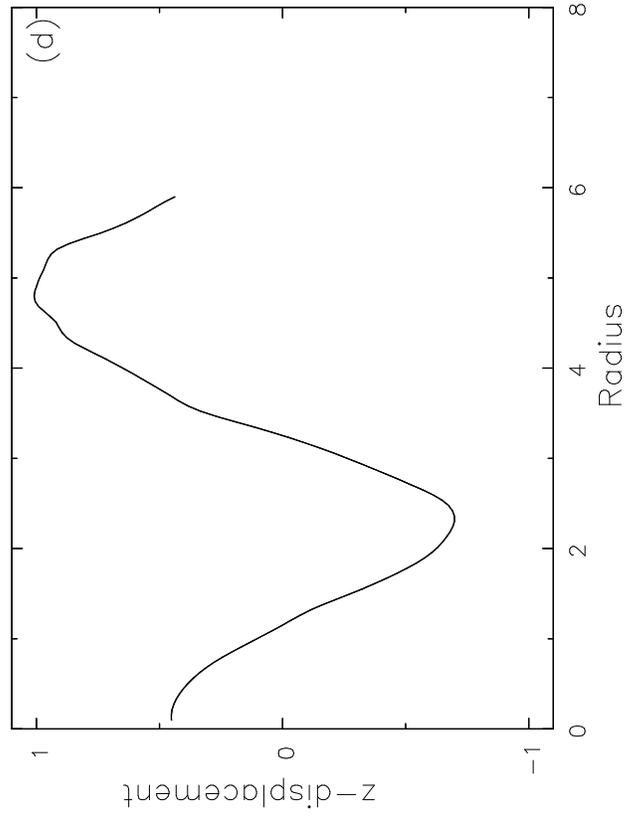
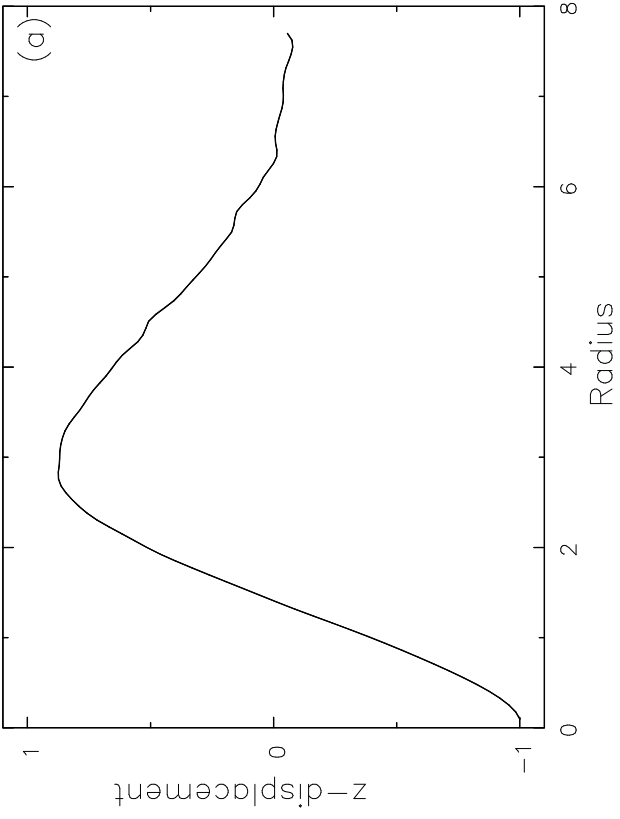
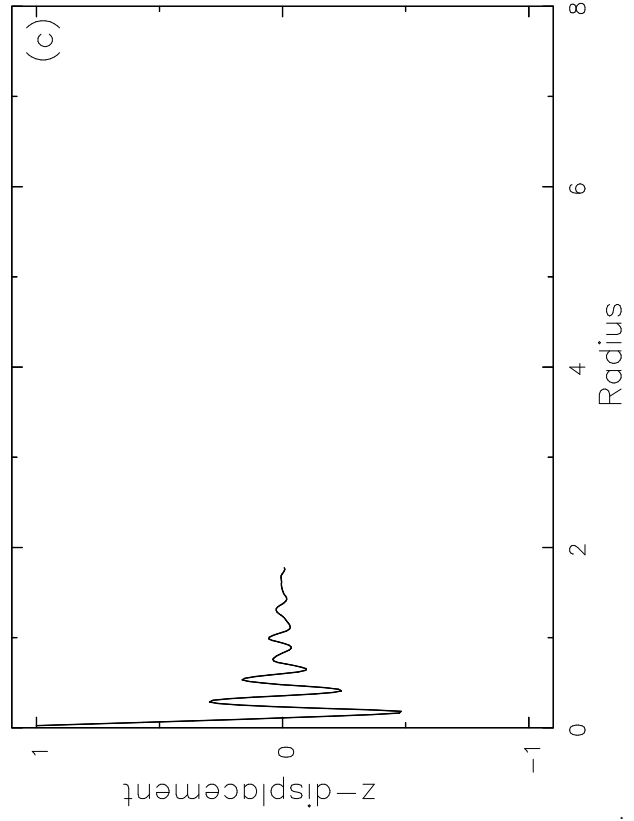
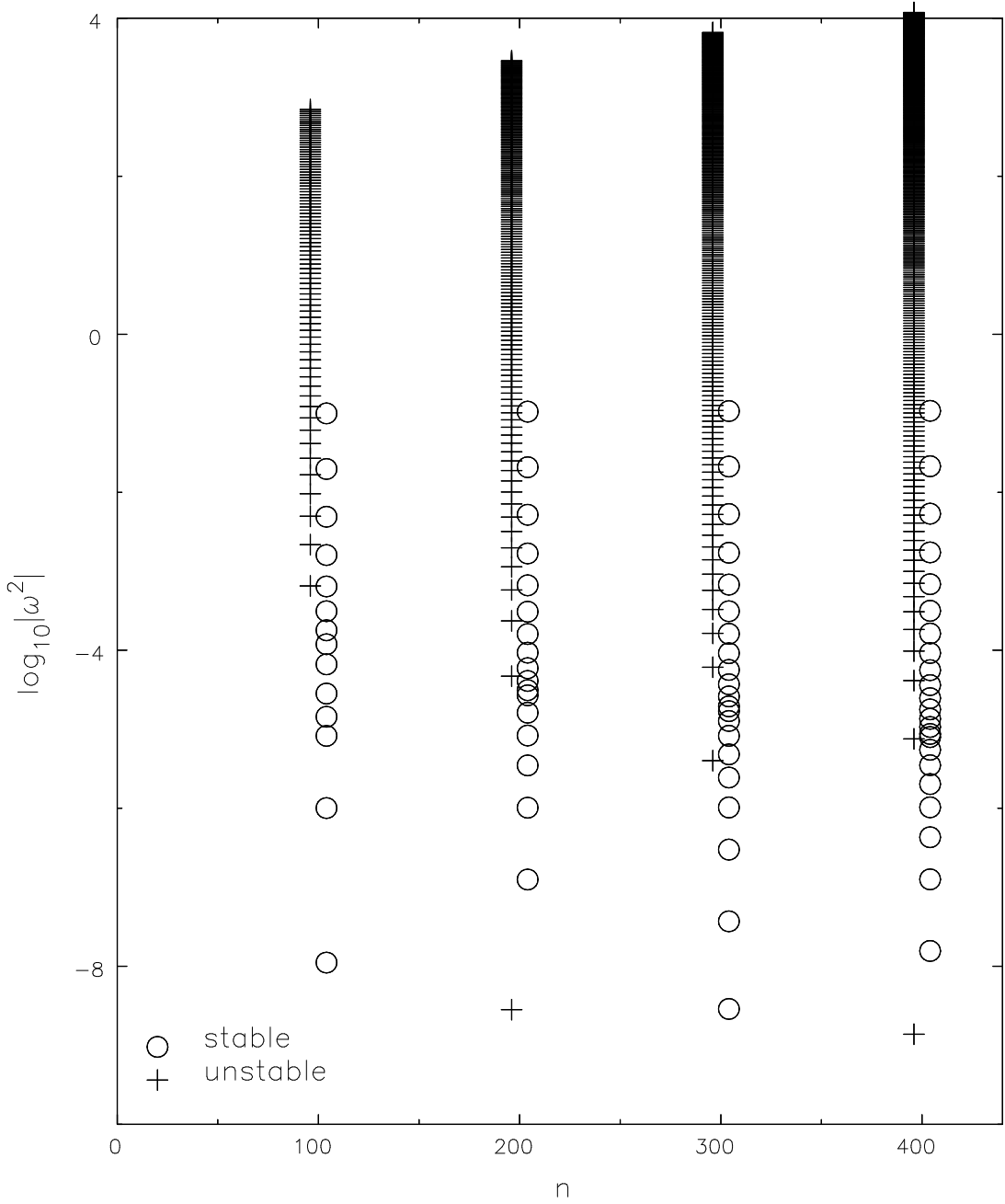
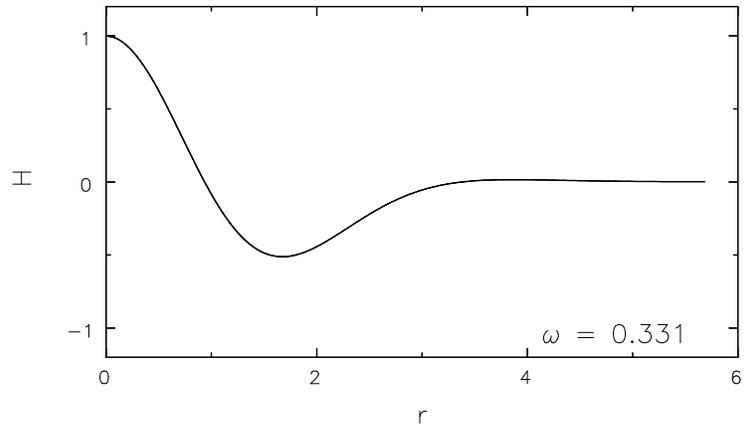
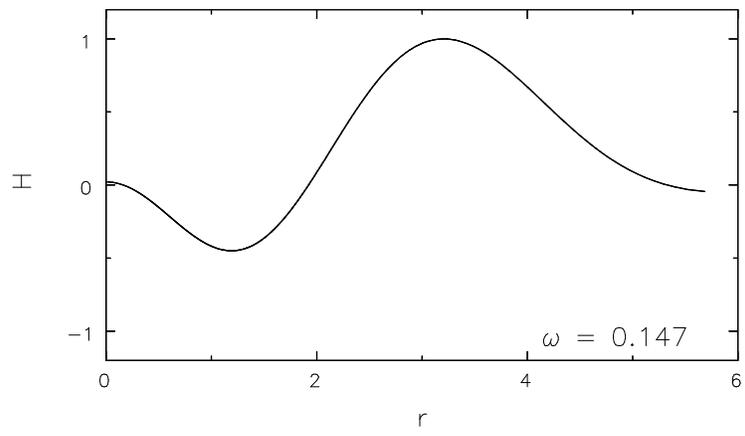
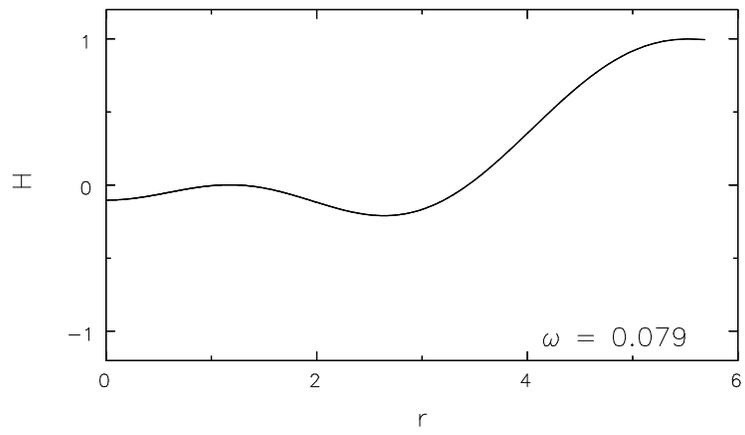
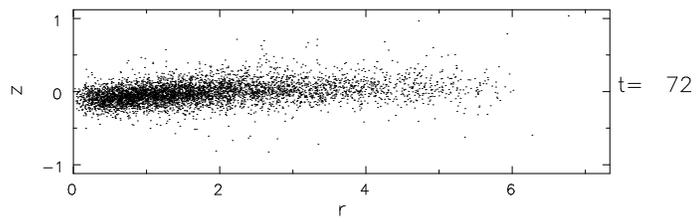
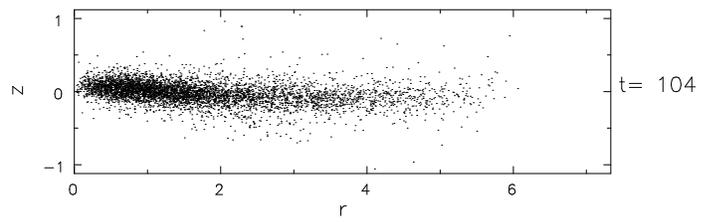
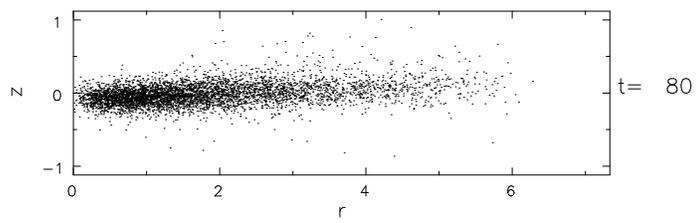
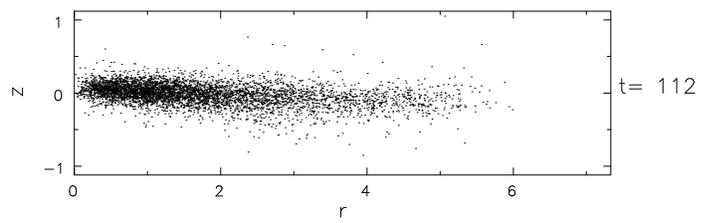
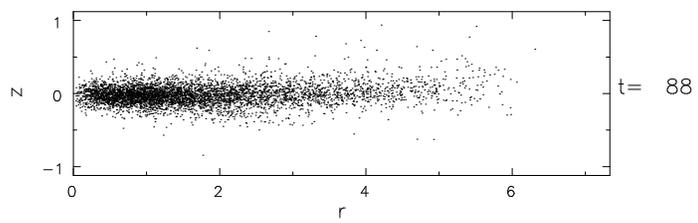
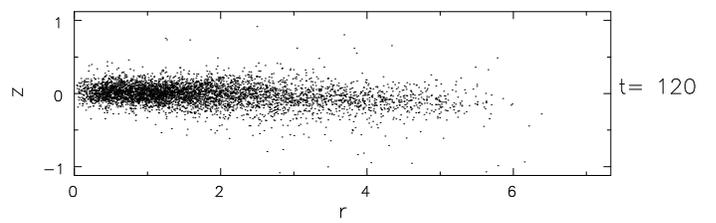
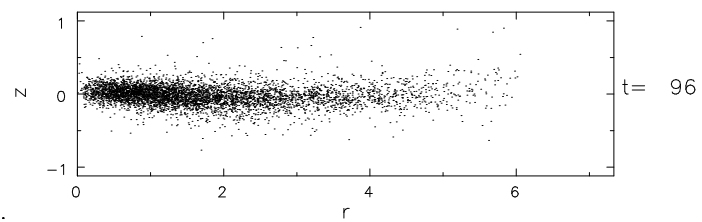
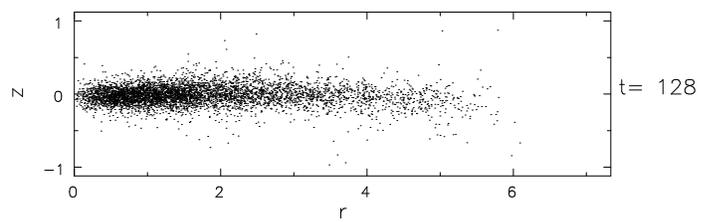
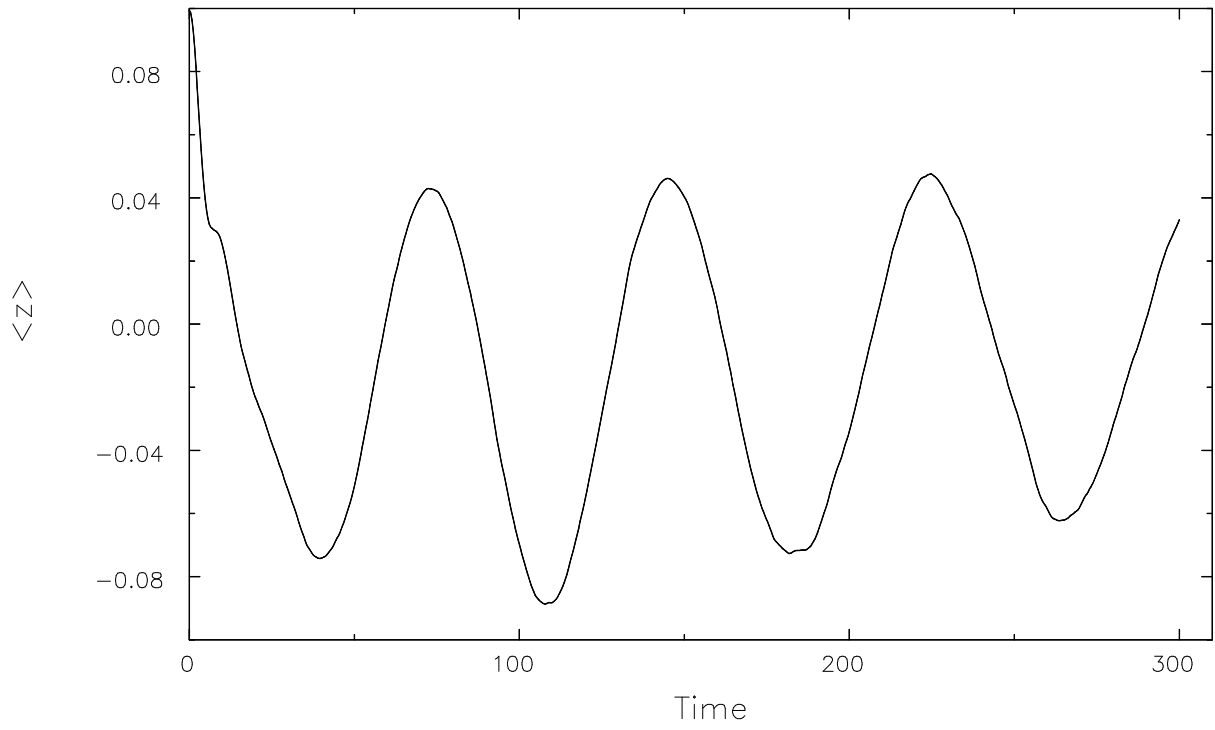
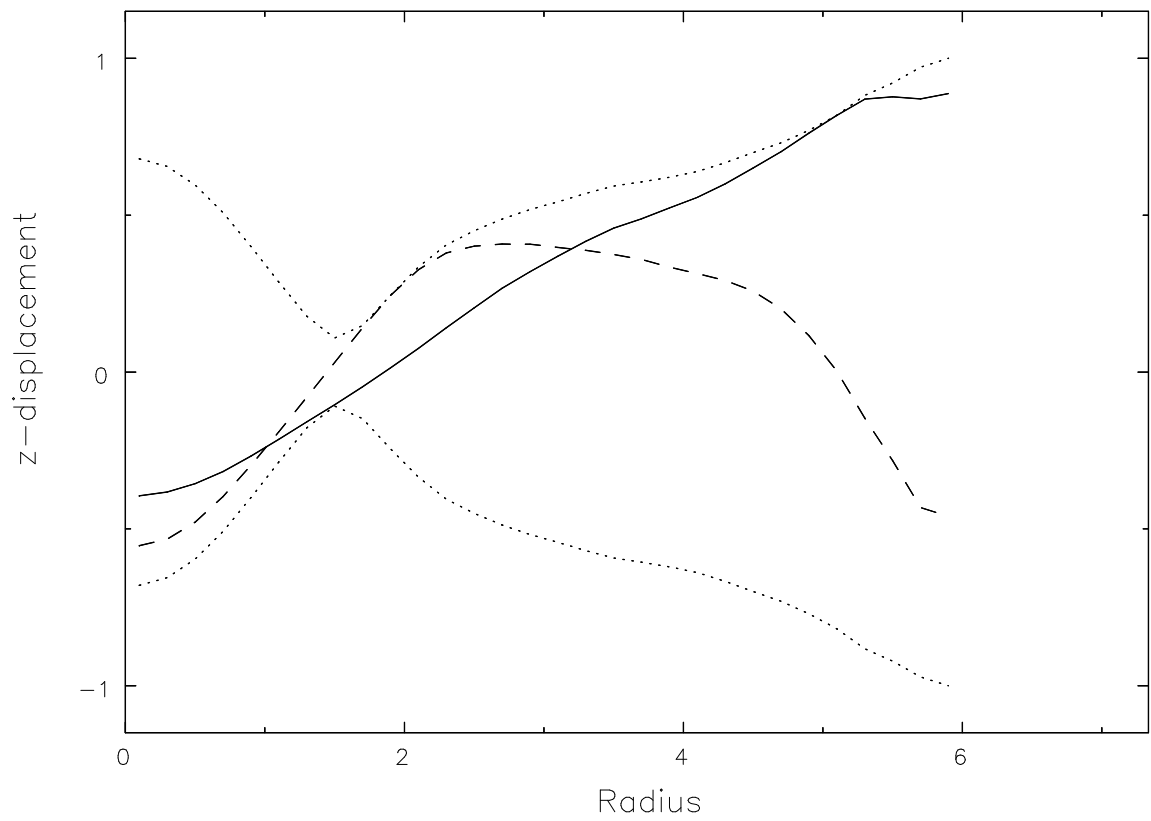
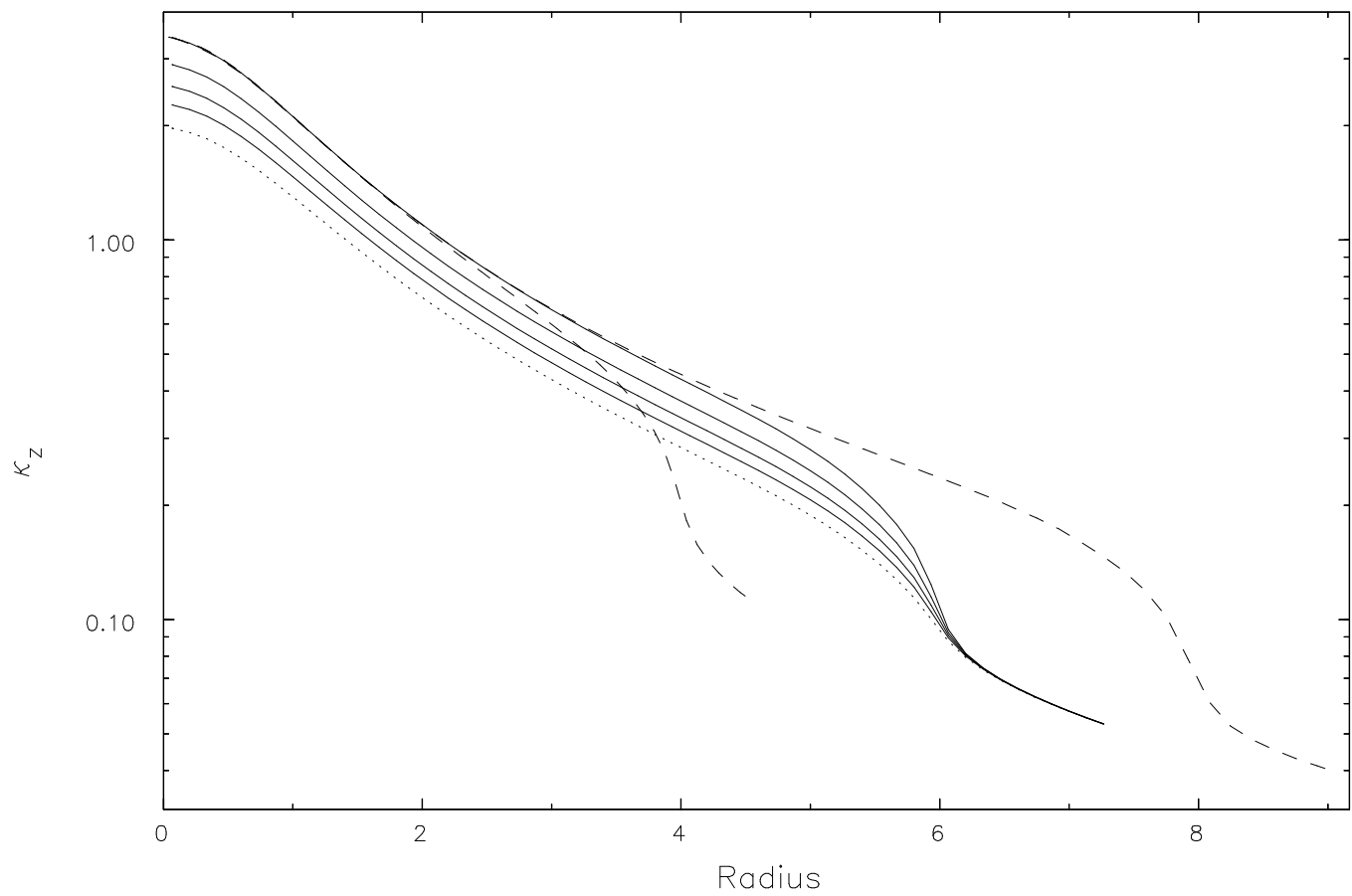
